\documentclass[pra,twocolumn,superscriptaddress,a4paper]{revtex4-2}
\usepackage{graphicx}
\usepackage{enumitem}
\usepackage{amsmath}
\usepackage{braket}
\usepackage{amsfonts}

\usepackage{xcolor}
\usepackage{amsmath}
\usepackage{amsfonts}
\usepackage{amssymb}
\usepackage{graphicx}
\usepackage{color}
\usepackage{bm}
\usepackage{soul}
\usepackage{subfigure}
\usepackage{changes}
\usepackage{comment}

\usepackage{array}
\newcolumntype{x}[1]{%
>{\centering\hspace{0pt}}p{#1}}%
\usepackage{multirow}

\usepackage[colorlinks,bookmarks=false,urlcolor=blue, citecolor=blue, linkcolor = blue,pdftitle={Impurity-induced pairing in two-dimensional Fermi gases},pdfborder={0 0 0},pdfborderstyle={}]{hyperref}

\DeclareMathOperator{\spn}{span}

\usepackage{layouts}
\usepackage[capitalise]{cleveref}
\begin{document}
\title{Impurity-induced pairing in two-dimensional Fermi gases}
\author{Ruipeng Li}
\altaffiliation{These authors contributed equally to this work}\email[E-Mail: ]{ruipeng.li@mpq.mpg.de, jvmilczewski@mpq.mpg.de}
\affiliation{Max-Planck-Institute of Quantum Optics, Hans-Kopfermann-Strasse 1, 85748 Garching, Germany}
\affiliation{Munich Center for Quantum Science and Technology (MCQST), Schellingstr. 4, 80799 Munich, Germany}
\author{Jonas von Milczewski}
\altaffiliation{These authors contributed equally to this work}\email[E-Mail: ]{ruipeng.li@mpq.mpg.de, jvmilczewski@mpq.mpg.de}
\affiliation{Max-Planck-Institute of Quantum Optics, Hans-Kopfermann-Strasse 1, 85748 Garching, Germany}
\affiliation{Munich Center for Quantum Science and Technology (MCQST), Schellingstr. 4, 80799 Munich, Germany}
\author{Atac Imamoglu}
\affiliation{Institute of Quantum Electronics ETH Zurich, CH-8093 Zurich, Switzerland}
\author{Rafa{\l} O{\l}dziejewski}
\email{rafal.oldziejewski@mpq.mpg.de}
\affiliation{Max-Planck-Institute of Quantum Optics, Hans-Kopfermann-Strasse 1, 85748 Garching, Germany}
\affiliation{Munich Center for Quantum Science and Technology (MCQST), Schellingstr. 4, 80799 Munich, Germany}
\author{Richard Schmidt}
\email{richard.schmidt@thphys.uni-heidelberg.de}
\affiliation{Max-Planck-Institute of Quantum Optics, Hans-Kopfermann-Strasse 1, 85748 Garching, Germany}
\affiliation{Institute for Theoretical Physics, Heidelberg University, Philosophenweg 16, 69120 Heidelberg, Germany}

\date{\today}

\begin{abstract}
We study induced pairing between two identical fermions mediated by an attractively interacting quantum impurity in  two-dimensional systems. Based on a Stochastic Variational Method (SVM), we investigate the influence of confinement and finite interaction range effects on the mass ratio beyond which the ground state of the quantum three-body problem undergoes a transition from a composite bosonic trimer to an unbound dimer-fermion state. We find that confinement as well as a finite interaction range can greatly enhance trimer stability, bringing it within reach of experimental implementations such as found in ultracold atom systems. In the context of solid-state physics, our solution of the confined three-body problem shows that exciton-mediated interactions can become so dominant that they can even overcome  detrimental Coulomb repulsion between electrons in atomically-thin semiconductors. Our work thus paves the way towards a universal understanding of boson-induced pairing across various fermionic systems at finite density, and opens perspectives towards realizing novel forms of electron pairing beyond the conventional paradigm of Cooper pair formation.
\end{abstract}

\maketitle

\section{Introduction}\label{sec:intro}

Frequently, the relevant physics of a many-body system is determined by the properties of its few-particle correlators, and thus a deep understanding of a many-body problem often comes only after carefully examining its few-body counterpart. An excellent example is given by the discovery of  Cooper pair formation as the key ingredient leading to superconductivity~\cite{Cooper1956,BCS1957}. No matter the type of a superconductor,  be it $s$-wave, $p$-wave, $d$-wave, or other like charge-$4e$ superconductors \cite{ketterson1999,Scalapino2012,Mackenzie2000,Viyuela2018,kirtley1995,Kivelson1990,Shao2021, Uchoa2007, nandkishore2012, baugher2014}, the phenomenon requires electrons to be bound into bosonic compounds. While, for conventional superconductors, the binding originates from phonon-mediated attraction, a variety of bosons ---partially stemming from collective excitations of the electronic system itself--- have been considered as the mediating particle \cite{Wheatley1975,Leggett1975,vollhardt2013superfluid,Stewart2017}.

More generally, quantum impurity-mediated pairing of fermions in the    mass-imbalanced $1+N$ fermion problem has been scrutinized extensively in recent years~\cite{Petrov2003,efimov1973,Kartavtsev2007, Stringari2008,Levinsen2009, Castin2010, Pricoupenko2010, Blume2012, massignan2014, Levinsen2015, Petrov2017, Petrov2018}. The vast majority of theoretical efforts have focused on non-interacting fermions and point-like impurity-fermion attraction that can be studied experimentally with ultracold gases~\cite{zwierlein2009,kohstall2012metastability,cetina2016ultrafast}. Interestingly, in the unconfined case, the system   supports cluster-bound states whenever the mass $m_I$ of the quantum impurity is sufficiently light compared to the mass $m_F$ of the fermions  . The critical mass ratio $\alpha = m_F/m_I$ required for such bound states to appear depends   on the dimensionality of the system: in  two dimensions (2D), the role of interactions is enhanced, and hence the mass ratio can be smaller compared to the three dimensional (3D) case ~\cite{Jesper2013,Cui2022}.  

A recent twist to the quantum impurity problem in 2D emerged with the advent of atomically-thin van der Waals materials, particularly semiconducting transition metal dichalcogenides (TMDs)~\cite{sidler2017fermi,wang2018tmdspectroscopy}. In TMDs,  excitons (bosons) can be either employed as an experimental probe of the many-body physics exhibited by electrons (fermions), ranging from Mott physics \cite{Shimazaki2021}, excitonic insulators \cite{Amelio2022} and the fractional Quantum Hall effect \cite{Popert2022} to the recent observation of Wigner crystallisation~\cite{smolenski2021,zhou2021}, or they can be viewed as novel constituents of Bose-Fermi mixtures \cite{sidler2017fermi,vonMilczewski2022,Imamoglu2021}, potentially supporting superconductivity~\cite{laussy2010,Cotlet2016, Kinnunen2018}. Importantly, in this case,  strong Coulomb repulsion is present between the fermionic electrons, and the impurity-fermion interaction itself is characterized by a substantial range~\cite{Fey2020}. So far, little is known about the existence and character of bosonic cluster-bound states in such a scenario.

\begin{figure*}[t!]
\centering
\includegraphics[width=\linewidth]{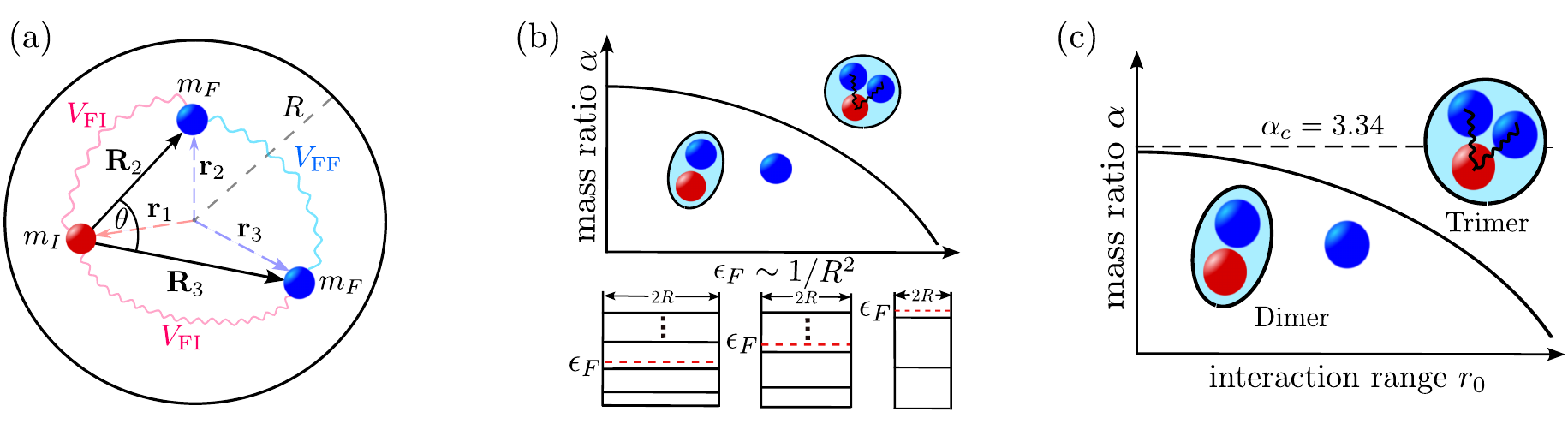}
\caption{Impurity-induced fermion pairing. (a) Illustration of the three-body system solved in this work. A mobile impurity particle (``I", red) of mass $m_I$ interacts with two fermions (``F", blue) of mass $m_F$ with all particles confined in a two-dimensional spherical box of radius $R$. Their positions, measured from the center of the box potential, are denoted by $\mathbf{r}_1$, $\mathbf{r}_2$ and $\mathbf{r}_3$. The coordinates $\mathbf{R}_2$, $\mathbf{R}_3$ and $\theta$, in turn, denote the positions of the fermions and their angle relative to the impurity, respectively. (b) Qualitative influence of the system size $R$ on the critical mass ratio $\alpha =m_F/m_I$ required for induced fermion pairing. By tuning the system size $R$, the density and Fermi energy $\epsilon_F$ of the fermions are tuned. This allows to infer how an increase of the Fermi level in a many-body system may enable impurity-induced bound state formation.  (c)
Qualitative effect of the  range $r_0$ of the impurity-fermion interaction on the critical mass ratio $\alpha_c =(m_F/m_I)_c$ of the dimer-trimer transition. The critical value  obtained in free space ($R \to \infty$) for contact interactions ($r_0 \to 0$) is shown as a dashed line. Both, increasing interaction range or the Fermi energy, favors trimer formation (which can even withstand detrimental Coulomb repulsion between the fermions, denoted as $V_{FF}$ in (a)).}\label{fig:system}
\end{figure*}

Recent advances in controlling 2D external confinement in
ultracold setups \cite{bloch2008,navon2021} and TMDs~\cite{xu2018} open an exciting possibility of exploring the physics of the quantum impurity problem in a fermionic background in a controlled bottom-up approach~\cite{bayha2020,holten2021,Holten2022}. Quite intriguingly, from the perspective of many-body physics, an alternative interpretation of the confinement potential is that of imitating a finite    fermion density found in many-body   paradigms such as  the Fermi polaron problem \cite{Schirotzek2009,Koschorreck2012,Ness2020,Fritsche2021}. Specifically, the change of the confinement ($\sim R$, see \cref{fig:system})  can be regarded as a primitive means of tuning the bath density ($n_F \sim 1/R^2\sim k_F^2$), realizing a few-body analogue of the full many-body problem \cite{Wenz2013}.

In this work, we significantly refine  previous understanding of   2D systems comprised of one impurity and two identical fermions (quantum statistically,  the smallest   Fermi sea possible) by studying the effects of a finite-range impurity-fermion potential, confinement, and strong inter-fermion repulsion on the ground state properties using a Stochastic Variational Method (SVM). As a key result, we show that the critical mass ratio  of the dimer-to-trimer transition   strongly departs from  previous findings obtained for the simpler case of ideal fermions and zero-range impurity-fermion attraction  (see \cref{fig:system}(b,c) for a schematic illustration). Remarkably, 
for  TMDs, where the transition occurs between a fermionic trion and a bosonic $p$-wave bound state of two electrons glued together by an exciton, we find that   trimer formation  is robust against Coulomb repulsion. Moreover, our numerical calculations show that the stability (in the sense of an increase of the dissociation energy required to unbind the trimer into a dimer state) of emerging bosonic $p$-wave bound states is enhanced by confinement. This suggests  that direct exciton-mediated $p$-wave superconductivity may be well in reach in solid-state systems.

\section{The model}\label{sec:model}
We consider an interacting system of two fermions and a quantum impurity confined in a two-dimensional spherical box; for an illustration, see \cref{fig:system}(a). This could represent two electrons interacting with an exciton in a quantum dot within a TMD, as well as two degenerate ultracold fermionic atoms interacting with an atom of a different quantum number within an oblate optical trap.  Using an effective mass approximation, the Hamiltonian for this system reads 
\begin{equation}
\begin{aligned}
H=&-\frac{\hbar^2}{2m_I}\nabla_1^2-\frac{\hbar^2}{2m_F}\nabla_2^2-\frac{\hbar^2}{2m_F}\nabla_3^2 +\sum_{i=1}^3 V_{\text{conf}} (\textbf{r}_i)\\ 
&+V_{\text{FI}}(\textbf{r}_1-\textbf{r}_2)
+V_{\text{FI}}(\textbf{r}_1-\textbf{r}_3)  + V_{\text{FF}}(\textbf{r}_2-\textbf{r}_3). \label{hamiltonian}
\end{aligned}
\end{equation}
Here $\textbf{r}_1$,  $\textbf{r}_2$ and $\textbf{r}_3$ denote the positions of the impurity and the two fermions, respectively, while $m_I$ and $m_F$ are their masses. The fourth term in the Hamiltonian represents the external confinement potential, which is modeled by an infinite potential well \footnote{In practice, this is achieved by setting $V_{\text{conf}} (\textbf{r})/E_{\text{ref}}= (\textbf{r}/R)^p  $ where $R$ is the box size, $E_{\text{ref}}$ is a reference energy scale   
and $p$ is a large integer so that an infinite potential well is approximated. In our calculation, we set $p=30$ and use the vacuum dimer energy as the reference energy $E_{\text{ref}}= E_{2B}^\infty$ .}. 

To account for finite range effects, the fermion-impurity interaction is modeled via a square well potential 
\begin{equation}
V_{\text{FI}}(\textbf{r})=
\left\{\begin{aligned}
& -V_0, && |\textbf{r}|\leq r_0 \\
& 0, && |\textbf{r}|> r_0
\end{aligned}\right. ,
\end{equation}
of depth $V_0$ and range $r_0$. Using this model potential, we mimic the finite range effects of the short-range interactions both in two-dimensional materials \cite{wang2018tmdspectroscopy,Fey2020} as well as ultracold atoms \cite{chin2010feshbach}.

A possible Coulomb interaction between the two fermions,
 \begin{align}
 V_{\text{FF}}(\textbf{r})&= \frac{e^2}{4\pi \epsilon_0 \epsilon}\frac{1}{|\mathbf{r}|} ,\label{eq:Coulomb}
 \end{align} 
is included by the last term in \cref{hamiltonian}. Here, $e$ is the electron charge and $\epsilon$ the dielectric constant of a given material. Note that in cold atoms this direct interaction is absent ($V_{\text{FF}}=0$). For TMD \cref{eq:Coulomb}  is a good approximation at large distance scales. At short range, the interaction between charge carriers is more accurately  modeled using the Rytova-Keldysh potential \cite{Rytova1967,Keldysh1979}. However, to capture the essential physics of the interplay of Coulomb repulsion, confinement, and electron-exciton attraction, we restrict ourselves to the use of the pure Coulomb potential in \cref{eq:Coulomb}. On the one hand, this allows for efficient numerics, and, on the other hand, this does not complicate the analysis by introducing  additional physical tuning parameters, such as the screening length. In the following, we set $\hbar=1$, unless  stated otherwise. 

\section{Method}

Apart from the task of solving the quantum mechanical problem of three interacting  particles, this system brings with itself the challenge of the additional confinement potential. This confinement is, however, crucial in order to imitate the effect of a finite fermion density $n_F$ in many-body systems, which scales as $n_F \sim 1/R^2\sim k_F^2$. Here $k_F$ denotes the Fermi wavevector of the fermions. The confinement breaks translational symmetry and thus is not susceptible to momentum space approaches using conventional variational wave functions or quantum field theory and diagrammatic   methods.

To solve for the ground state and its energy, we employ the SVM \cite{suzuki1998}. To this end, the Hamiltonian $H$ is diagonalized with respect to a set of wavefunctions $\left \{ \Phi_n \right\}_{n=1}^N$ which is successively extended by drawing from a manifold of trial functions. In every extension step $N \to N+1$, the choice of the new wavefunction $\Phi_{N+1}$ is optimized in a stochastic random walk, minimizing the lowest-lying eigenstates of the Hamiltonian $H$ with respect to the vector space spanned by the set $\left \{ \Phi_n \right\}_{n=1}^{N+1}$. During the optimization, we first draw a set of independent samples from the manifold of trial functions and then perform a random descend walk around the best proposal state. Having performed an extension step, the Hamiltonian $H$ is diagonalized with respect to the vector space spanned by the $\left \{ \Phi_n \right\}_{n=1}^{N+1}$.  The resulting $i$-lowest eigenstate $\Psi_{i}$ is then given by a superposition of these basis states, i.e.    $\Psi_{i}= \sum_{n=1}^{N+1} c_n^i \Phi_n$, where $i=1,...,N+1$ and the eigenstates $\left \{ \Psi_i \right\}_{i=1}^{N+1}$ are mutually orthogonal.

In many applications of SVM, trial functions are generated from explicitly correlated Gaussians (ECG). These are, parametrized as $\Phi(\mathbf{r}_1,\mathbf{r}_2,\mathbf{r}_3)=\mathcal{P}\exp{\left(-\frac{1}{2}\sum_{i,j=1}^3A_{ij}\mathbf{r}_i\cdot\mathbf{r}_j\right)} /r_B^3$ \cite{suzuki1998, Mitroy2013}.  Here, $A$  denotes a positive definite, symmetric $3\times 3$ matrix, $\mathcal{P}$ is an antisymmetrization operator.  The length scale $r_B$,  introduced in \cref{sect:ground_state},   characterizes the size of the dimer bound state. The advantage of using these trial functions is threefold. First, they allow one to find the analytical solution to the matrix elements of the Hamiltonian  \cite{Varga2008,Mitroy2013}. Second, by using them, high accuracy in the energy can be achieved. Finally,  the ECG contain the relevant physical states (dimers, trimers, and scattering states in our system) and, as such, they have been used to calculate exciton, trion and even biexciton energies in solid state systems with high precision~\cite{Cho2021,Yan2020,Donck2018,Kidd2016}. For more detail on the optimization algorithm, the sampling from the ECG manifold and the computation of expectation values with respect to the ECG manifold, we refer to \cref{svmalgorithm}.

\section{Ground state}\label{sect:ground_state}

In this section, we calculate the ground state using the SVM. As the 2D system features binding via the fermion-impurity potential $V_{\text{FI}}$ for any potential depth \footnote{As long as the size of the impurity-fermion bound state is smaller than the confinement length scale.}, states composed of a dimer and a  fermion in a scattering state are expected to play a vital role \cite{Adhikari1986}. Moreover,  for sufficiently light impurities, the formation of a trimer is expected. In this state, two fermions and the impurity bind together by the mediating force of the impurity \cite{Petrov2018}. This is similar to the three-dimensional case  where a $p$-wave trimer and eventually Efimov states appear for sufficiently light impurities \cite{Petrov2003,Kartavtsev2007,Levinsen2009}.  

In the limit of a vanishing interaction range $r_0 \to 0$ and infinite system size $R\to \infty$, a ground state transition from a dimer to a trimer state is predicted to occur when the mass ratio $\alpha=m_F/m_I$ is tuned across the critical value $\alpha_{c}\approx3.34$ \cite{Pricoupenko2010,Jesper2013, Cui2022}. Having this  limiting case  as a benchmark,  we investigate the effect of interaction range $r_0$ and confinement (determined by the system size $R$) on  the critical value $\alpha_{c}$. It is important to note that the transition will occur as a crossover because of the finite size of the system. Specifically, we study how the ground state characteristics and energy   change as we tune $\alpha$, $r_0$, and $R$  and, as a result, how the critical mass ratio varies with $r_0$ and $R$. 

In the following, we will refer to the two-body bound state appearing in an untrapped ($R\to \infty$) two-body problem consisting of the impurity and a fermion as the `vacuum dimer'.  Its binding energy will be denoted as the `vacuum dimer energy' $E_{2B}^{\infty}$. The terms `dimer' and `trimer', in turn, will refer to states in the three-body problem. Specifically,  the  dimer refers to a state comprised of a fermion in a scattering state along with a two-body bound state of an impurity and a fermion, while the  trimer denotes a three-body bound state consisting of an impurity bound to  both fermions.  

\begin{figure*}[t!]
\centering
\includegraphics[width=\linewidth]{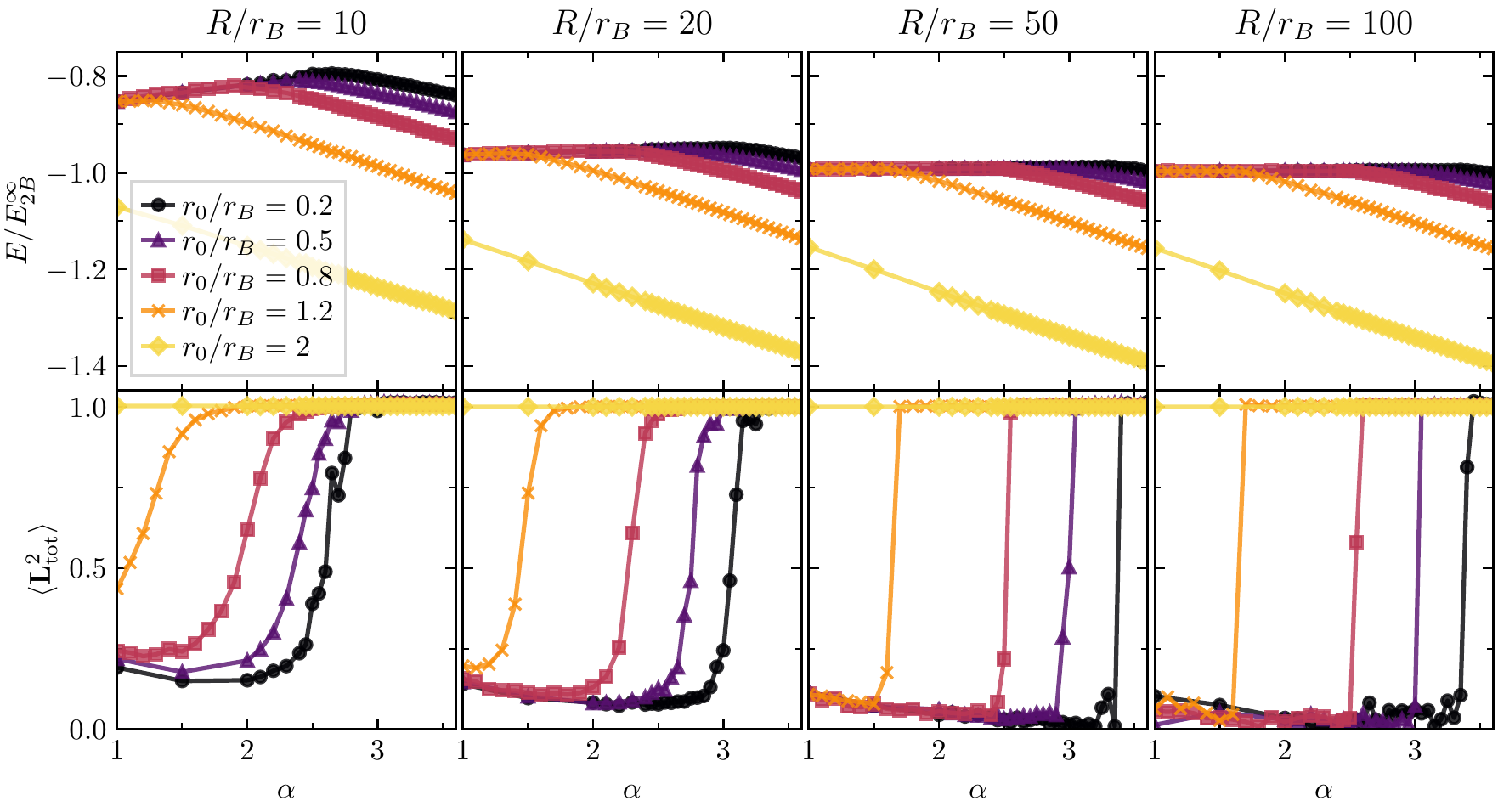}
\caption{Energies $E/E^{\infty}_{2B}$ (top panels) and angular momentum expectation values $\braket{\mathbf{L}^2_{\text{tot}}}$ (bottom panels) of the ground state as a function of mass imbalance $\alpha$ for various dimensionless interaction ranges $r_0/r_B$ and system sizes $R/r_B$. The energies are located around $E^{\infty}_{2B}$ with upward shifts mainly due to the confinement contributions to the  kinetic energy of the particles. The crossover from the dimer to the $p$-wave trimer bound state is  visible in the angular momentum (lower panels) which crosses over from being close to $0$ to approximately $1$.  This crossover is similarly reflected in a drop of the ground state energy  which develops an almost linear dependence on $\alpha$ beyond the crossover from the  dimer to the trimer ground state. For increasing values of $r_0/r_{B}$ ($R/r_{B}$),  this crossover region  is shifted to lower (higher) mass ratios $\alpha$. For $R/r_{B}\to \infty$, the crossover becomes a sharp transition which, for $r_0/r_{B}\to 0 $, occurs at $\alpha_{c}\approx 3.34$ \cite{Pricoupenko2010,Jesper2013,Cui2022}.}
\label{energyangular}
\end{figure*}

To study the dimer-to-trimer transition,  we vary $\alpha$, $r_0$ and $R$ while keeping the non-trapped ($R\to\infty$) vacuum dimer energy $E_{2B}^{\infty}$ constant.  We define a corresponding binding length  $r_B= 1/\sqrt{2 m_F E_{2B}^{\infty}}$, and, unless explicitly stated otherwise, we will work in units where the fermion mass is set to $m_F=1/2$. Note,  that we have  defined $r_B$ by the fermion mass and not the reduced mass. This convention  ensures a fixed value of $r_B$ as $\alpha$ is changed. One has to keep in mind, however, that now $r_B$ is  proportional to the physical binding length of the dimer state. The two-body Hamiltonian of one fermion and one impurity interacting via  $V_{\text{FI}}$ can be solved exactly \cite{Whitehead2016}. As detailed in \cref{twobodyall}, this allows to obtain the required potential depth $V_0$ for  given values of  $\alpha$, $r_0$, and  $E_{2B}^{\infty}$.

We now begin our numerical study by first considering the system without Coulomb interactions ($V_{\text{FF}}=0$). After establishing the dimer-to-trimer transition for this case, we will switch on Coulomb interactions ($V_{\text{FF}}> 0$), and systematically explore their effect.

\subsection{Non-interacting fermions}
\label{nonintferm}

\begin{figure}[t!]
\centering
\includegraphics[width=\linewidth]{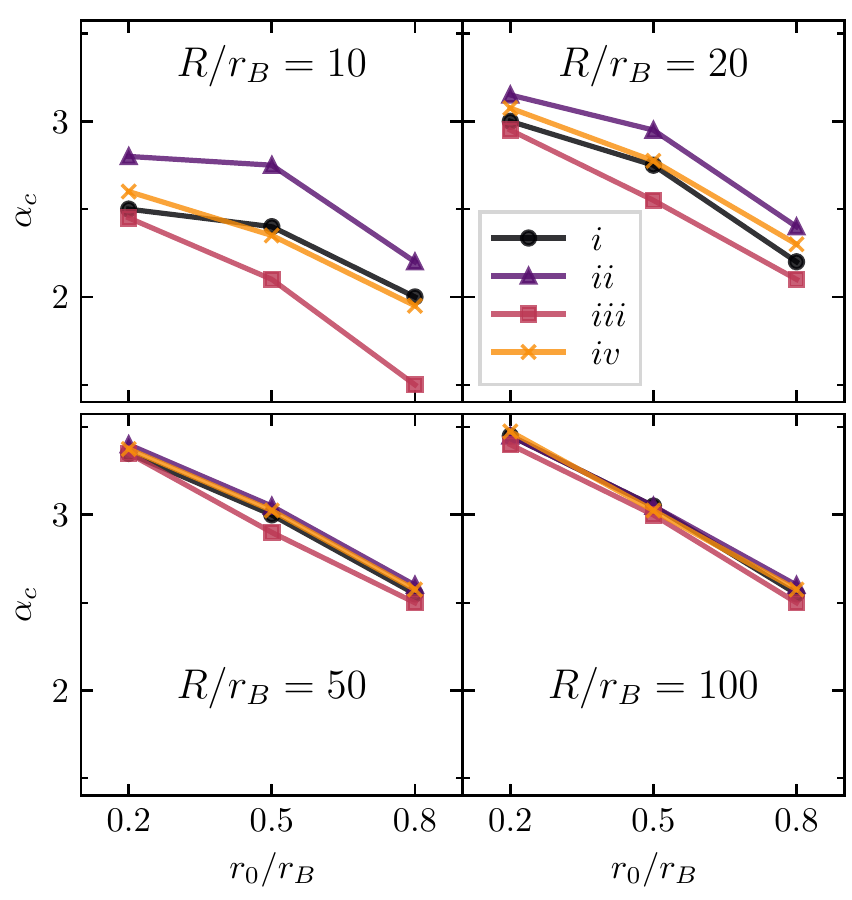}
\caption{Critical mass ratio for the dimer-to-trimer transition as function of the interaction range  $r_0/r_B$ for $R/r_B=10$, $20$, $50$, $100$. The mass ratios are determined using four different criteria: \romannumeral1.  appearance of ground state energy decrease (black dots); \romannumeral2. $\langle\mathbf{L}_{\text{tot}}^2\rangle\approx1$ (purple triangles); \romannumeral3. $\langle\mathbf{L}_{\text{tot}}^2\rangle\approx0$ (red squares); \romannumeral4. $\langle\mathbf{L}_{\text{tot}}^2\rangle\approx0.5$ (yellow crosses). The different criteria  lead to different values of $\alpha_c$, with the $\langle\mathbf{L}_{\text{tot}}^2\rangle\approx0$ criterion consistently giving the lowest mass ratio, while the $\langle\mathbf{L}_{\text{tot}}^2\rangle\approx1$ criterion yields the highest. With increasing $R/r_{B}$, the crossover region becomes more narrow, and the results from the different methods converge.}
\label{alphacrit}
\end{figure}

In \cref{energyangular}, we show the energy of the SVM ground state as function of the mass ratio $\alpha$, for different values of $r_0$ and $R$. Here, $r_0$ and $R$ are varied in terms of the dimensionless quantities $r_0/r_{B}$ and $R/r_{B}$. The ground state energies are all located in the vicinity of $-E^{\infty}_{2B}$. 
For fixed $r_0/r_{B}$ and $R/r_{B}$, the ground state energies first increase slightly with the mass ratio and then show a drop at  a critical mass ratio. Beyond the critical mass ratio, the ground state energy decreases steadily, exhibiting an almost linear dependence on the mass ratio, $E\propto - \alpha$ \cite{Pricoupenko2010,Jesper2013,Cui2022}. One can see that $r_0$ and $R$ have a strong influence  on the energies and the critical mass ratio at which the qualitative change in the ground state energy occurs. For a fixed system size $R/r_{B}$, upon increasing $r_0/r_{B}$, both the ground state  energies and the critical mass ratio  decrease. On the other hand, for a fixed interaction range $r_0/r_{B}$, an increase in system size $R$ leads to a decrease of the energy that is accompanied by an increase of the critical mass ratio.

We now turn  to a detailed discussion of the qualitative change observed in the ground state energy. This change signifies a transition of the ground state, where, for values of $\alpha$ smaller than a critical value, the system is in the `dimer' state, i.e. it is composed of a bound dimer along with an unbound fermion. In contrast, beyond the critical value of $\alpha$,  the ground state energy falls below the  dimer-fermion scattering threshold  energy,  indicating the emergence of the trimer state,  similar to the unconfined system \cite{Pricoupenko2010,Jesper2013,Cui2022}.

While the energy is a good indicator of a qualitative change,  a reliable identification of  the nature of the ground state  requires a  deeper analysis of the corresponding wave function. In the following, we will  show that the angular momentum and  the density distribution provide two  measures to clearly distinguish the dimer and trimer state. 

First, we focus on the analysis of  angular momentum. To this end, we introduce the relative coordinates  $\mathbf{R}_2=\mathbf{r}_2-\mathbf{r}_1$ and $\mathbf{R}_3=\mathbf{r}_3-\mathbf{r}_1$, where $\mathbf{R}_2$ and $\mathbf{R}_3$ denote the  positions of the fermions relative to the impurity. The total angular momentum relative to the impurity particle is then given by $\mathbf{L}_{\text{tot}}=\mathbf{L}_2+\mathbf{L}_3$, where $\mathbf{L}_2=\mathbf{R}_2\times\mathbf{P}_2$ and $\mathbf{L}_3=\mathbf{R}_3\times\mathbf{P}_3$. Here,  $\mathbf{P}_2$ and $\mathbf{P}_3$ are the momentum operators corresponding to $\mathbf{R}_2$ and $\mathbf{R}_3$, respectively.  In this relative coordinate frame,  fermionic statistics imposes the trimer  to have odd, finite angular momentum $\braket{\mathbf{L}_{\text{tot}}}=\pm1$, while the dimer state has $\braket{\mathbf{L}_{\text{tot}}}=0$ \cite{Pricoupenko2010,Becker2018,Jesper2013,Petrov2018,Cui2022}.

As a result of the ECG functions we use, the basis functions are real and hence any measured value of $\braket{\mathbf{L}_{\text{tot}}}$ has to vanish. As a consequence of this constraint,  the  wavefunction of the trimer state  obtained from the SVM is an equal superposition of degenerate ground states with $\braket{\mathbf{L}_{\text{tot}}}=1$ and $\braket{\mathbf{L}_{\text{tot}}}=-1$; resulting in the expectation value $\braket{\mathbf{L}_{\text{tot}}}=0$. Thus, in order to obtain a  characterization of the ground state, we consider the expectation value 
$\braket{\mathbf{L}^2_{\text{tot}}}$. This  allows us to distinguish the dimer and trimer state in a reliable way (for more details, we refer to \cref{svmalgorithm}). 

We show the ground state value of $\braket{\mathbf{L}^2_{\text{tot}}}$ in the lower column of
\cref{energyangular}. As one can see, $\braket{\mathbf{L}^2_{\text{tot}}}$ sharply increases from values close to 0 to approximately 1 as the mass ratio is tuned beyond a critical value. The region in which this qualitative change occurs coincides with the critical mass ratio at which the drop in energy is observed (upper panels of \cref{energyangular}). The close link between the behaviour of the ground state energy and angular momentum is robust across all values of  $r_0/r_{B}$ and $R/r_{B}$.  
While for smaller system sizes, the transition region is larger,  with increasing system size, the transition region becomes more narrow. This indicates that, as expected,  the crossover found for a finite system turns into a sharp transition for an infinite system size.

From the behavior of energy and angular momentum, a simple physical picture of the crossover from a dimer to a trimer arises. At smaller mass ratios $\alpha$, the ground state is given by a dimer along with a fermion in a delocalized scattering state. Thus, for large system sizes, the energy approaches the two-body energy $-E^{\infty}_{2B}$. However, for smaller system sizes the confinement induces \mbox{exchange-,} correlation- and confinement-energies between the two fermions increasing the energy above $-E^{\infty}_{2B}$. This increase in energy is larger for smaller system sizes and features an additional weak dependence on the mass ratio that can be understood already from the non-interacting system where the confinement energy is given by $E_{\text{conf}}=z_{01}^2/2m_IR^2+z_{11}^2/m_FR^2=(z_{01}^2\alpha/2+z_{11}^2)/m_FR^2$ with $z_{01}$ and $z_{11}$  the first 
zeros of the Bessel functions $J_0$ and $J_1$, respectively. Beyond the critical mass ratio, the ground state is described by a trimer state, and its energy starts to decrease close to linearly with the mass ratio, as also found  in the  continuum case \cite{Pricoupenko2010,Jesper2013,Cui2022}.

 We now turn to a more detailed analysis of how the system size $R$ and interaction range $r_0$ affect the critical mass ratio $\alpha_c$ (see  \cref{alphacrit}).  Decreasing the system size has a stronger effect on the dimer state than on the trimer state. This is caused by the fact that   the unbound fermion in its delocalized scattering state feels the confinement more strongly than a fermion bound tightly to the impurity.  As a result, the trimer state is subject to a confinement energy contribution less than the dimer state. Consequently, decreasing system size moves the transition to smaller mass ratios.

Increasing the interaction range $r_0$ affects the trimer state stronger than it affects the dimer state.  For $R\gg r_0$, the average distance between  the fermions in a trimer state is related to the short distance scales $r_B$ and $r_0$ while, in the dimer state 
(which includes the unbound fermion), it is related to $R$. Thus, increasing $r_0$, lowers the Pauli-repulsion  within the trimer state, making the trimer favorable which,  decreases the critical mass ratio.  This intuitive picture is reflected in the numerical results presented in  \cref{alphacrit}. In this Figure, we additionally analyze the increasing sharpness of the transition as the system size is increased by showing the critical mass ratio as obtained from different criteria imposed on the energy and the angular momentum. As one can see, for $R/r_B=100$,  all  criteria give nearly identical results, and only the dependence on the scale $r_0$ remains. 

As can be seen from the lower panel in \cref{energyangular}, the impact of the interaction range and  system size on the dimer and trimer state is also reflected in the angular momentum. Due to the confinement, the free fermion in the dimer state is forced to take on a finite angular momentum state, resulting in a nonzero value of $\braket{\mathbf{L}^2_{\text{tot}}}$. As the system size is increased, the free fermion is less affected and $\braket{\mathbf{L}^2_{\text{tot}}}$ approaches zero. The trimer, on the other hand,  is hardly affected by the finite system size as long as $R\gg r_0$, and thus $\braket{\mathbf{L}^2_{\text{tot}}}$ is very close to $1$. 

Our finding of a strong dependence of $\alpha_c$ on  $r_0$ and $R$ shows that the critical mass ratio of 3.34, obtained in the limit $r_0\to 0$ and $R\to \infty$ \cite{Pricoupenko2010,Jesper2013,Cui2022}, potentially features only a small window of universality. In this regard we note that the critical mass ratios in \cref{alphacrit} for $R/r_B=100$, $r_0/r_B=0.2$ tend to lie slightly higher than the asymptotic value of 3.34. This is due to the stochastic nature of our method which is particularly challenged when the energetic difference between dimer and trimer particles becomes very small, which precisely occurs close to the transition.  As a result, especially for larger system size and shorter interaction range, a suitable trimer wavefunction can only be found for a high number of proposed wave functions.  In \cref{app:convergence_analysis}, the deviation from the asymptotic value of $\alpha_c=3.34$ is studied in detail, and additionally, a convergence analysis, including an estimate for the basis set extrapolation error, is undertaken. For further details on the sampling methods used in this work, we refer also to \cref{svmalgorithm}.

\begin{figure}[t!]
\centering
\includegraphics[width=\linewidth]{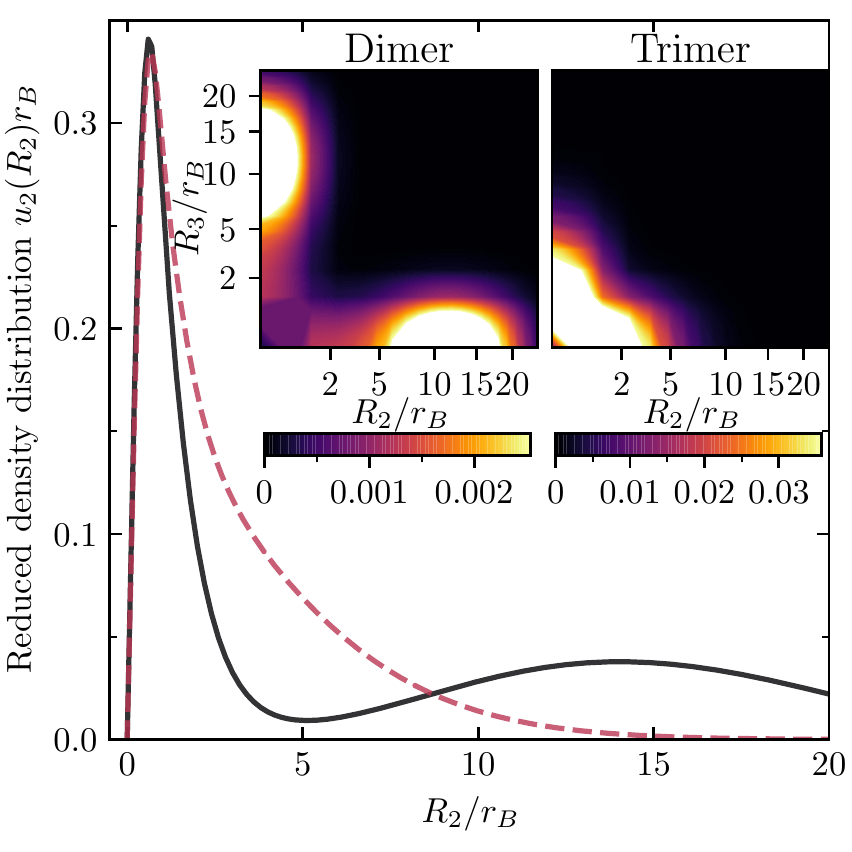}
\caption{Reduced density distributions of a dimer ($\alpha=2$) and trimer ($\alpha=3$) state. The main plot shows $u_2(R_2)r_B$ for the dimer (black, solid) and the trimer (red, dashed) state for $R/r_B=20$ and $r_0/r_B=0.8$. The exponential decay of the trimer distribution is clearly visible while the dimer state contains a fermion that is delocalized at the length scale of the system size. The  insets show $u_1(R_2,R_3)r_B^4$ for the dimer (left) and the trimer state (right).  For the trimer state, $u_1(R_2,R_3)r_B^4$ attains its largest values when $R_2$ and $R_3$ are both small, which shows that both fermions are close to the impurity, while for the dimer state $u_1(R_2,R_3)r_B^4$ attains its maximum on the x- and y-axis.}\label{densitydist}
\end{figure}

The spatial localization of the fermions around the impurity ---or the lack thereof--- provides a further means to confirm the presence of two- and three-body bound states. To that end, we study the spatial structure of the ground state wavefunction. It is expected that in the trimer state the two fermions are both close to the impurity, while in the dimer state, one fermion should be close to the impurity while the other resides in a delocalized scattering state. To study this behavior, we consider the correlation functions (which can be regarded as  reduced density distributions) 
\begin{align}
 u_1(R_2,R_3)&=\int |\Psi(\mathbf{r}_1,\mathbf{r}_1+\mathbf{R}_2,\mathbf{r}_1+\mathbf{R}_3)|^2d^2\mathbf{r}_1 d\theta_2 d\theta_3, \\
 u_2(R_2)&=R_2 \int dR_3  R_3 u_1(R_2,R_3).
\end{align}
Here, $\Psi$ denotes the three-body wave function, and the angles $\theta_2$, $\theta_3$ are defined via $\mathbf{R}_2=R_2 (\cos\theta_2,\sin \theta_2)$ and $\mathbf{R}_3=R_3 (\cos\theta_3,\sin \theta_3)$. From this definition, one can see that the reduced density distribution $u_1$ measures the probability of simultaneously finding one electron at a distance $R_2$ while  the other is situated at distance $R_3$ from the impurity. The distribution is obtained by integrating out the coordinates of the impurity followed by a further  average over the angular orientation of the fermions with respect to the impurity.
Performing an additional integral over the distance of one of the fermions from the impurity, one  obtains a measure for the probability ($u_2$) of finding one fermion at a distance $R_2$ from the impurity. 

In \cref{densitydist},  density distributions are shown for exemplary trimer and dimer states. For the trimer state, the density distribution $u_2$ indeed exhibits an exponential decay, in line with the expectation that both fermions are closely-bound to the impurity. In contrast, for the dimer state, $u_2$ does not decay exponentially  but features a tail that corresponds to one of the fermions being situated in a scattering state. Note that for the confinement length of $R/r_B=20$ chosen in this figure, the distance between particles can be up to twice as large. Thus the density distribution does not vanish beyond $R_2/r_B=20$ but rather beyond the maximal interparticle distance (not shown in the graph).

Density plots  of the correlation function $u_1$ are shown in the inset of \cref{densitydist}. They give further insight into the anatomy of the dimer and trimer states with respect to their radial distribution. For the dimer state, the density distribution $u_1$ almost vanishes along the diagonal and achieves its maximum at approximately $(R_2/r_B, R_3/r_B)\approx (0,12)$. This exemplifies how in the dimer state one fermion is closely bound to the impurity while the other fermion is delocalized. For the trimer state, $u_1$ attains its largest values when $R_2$ and $R_3$ are both small. Moreover, $u_1$ vanishes rapidly for larger $R_2$ and $R_3$, which shows that both fermions are tightly bound to the impurity. However, the analysis of $u_1$ also reveals that, within the trimer state, there is always one fermion that is bound tightly  to the impurity, while the second fermion will be in a bound `orbit' at a slightly larger distance. In \cref{apptheta}, we analyze the angular configuration of both states and show that fermions tend to be located on opposite sides relative to the impurity.

\subsection{Coulomb interaction}
\begin{figure}[t!]
\centering
\includegraphics[width=\linewidth]{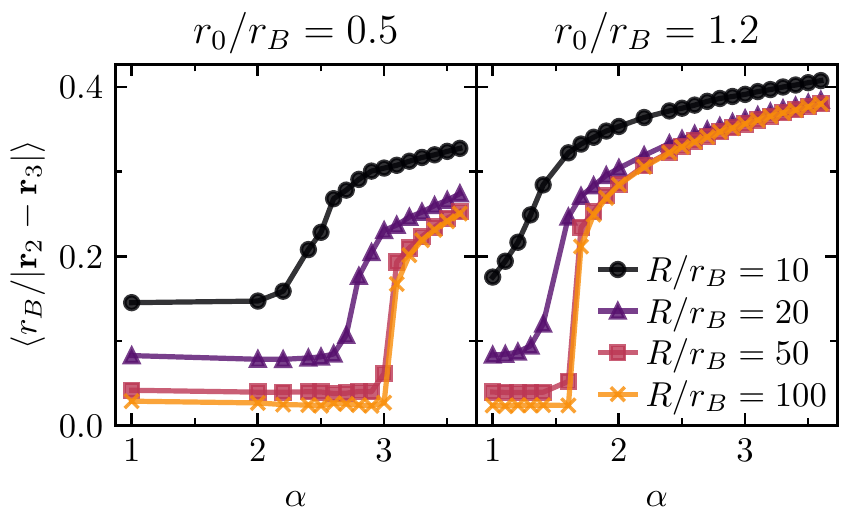}
\caption{Expectation value of $r_{B}/|\mathbf{r_2}-\mathbf{r_3}|$ of the ground state wavefunctions obtained in 
\cref{nonintferm}, for $r_0/r_{B}=0.5$ (left) and $1.2$ (right), shown for a range of system sizes $R/r_B$. The crossover from the dimer to trimer state is visible in the steep increase of the expectation value. Increasing the box size $R$ and decreasing the interaction range $r_0$  moves the crossover to higher $\alpha$, consistent with our previous results. Moreover, the expectation value increases as the box size becomes smaller, because the confinement of the fermions in a smaller area   results in a larger Coulomb energy.}\label{fig:oneoverr}
\end{figure}

We now consider the impact a repulsive interaction potential between the two fermions ($V_{\mathrm{FF}} > 0$) has on the dimer-to-trimer transition. In particular, we focus on  Coulomb interactions present in 2D semiconductors (see \cref{eq:Coulomb}). In the trimer state, both electrons bind to the exciton bringing themselves closer together. Intuitively, this  can give rise to a considerable increase in the total energy of the cluster, weakening its binding. Consequently, given a fixed mass ratio, if the repulsive Coulomb energy becomes larger than the energy gap between the trimer and dimer states, the ground state is  expected to unbind into a state comprised of a dimer and a free electron.

To roughly estimate  the impact of the Coulomb energy on the total energy, we first calculate the expectation value of the Coulomb interaction  $\sim \braket{r_{B}/|\mathbf{r_2}-\mathbf{r_3}|}$ with respect to the ground state of the system without Fermi-Fermi interaction. We stress again~(see \cref{sec:model}) that, in the following, we shall use the Coulomb potential instead of a more accurate approximation of 2D interactions between charges given by the Keldysh potential. In any case, since the Coulomb interaction is more extreme than the Keldysh potential at  short range, we  expect our choice to be more restrictive than the Keldysh interaction (at short distance the Coulomb interaction diverges as $1/r$, while the Keldysh potential diverges as $\log(r/r_{sc})$; with $r_{sc}$ the screening length).

The expectation value of $r_{B}/|\mathbf{r_2}-\mathbf{r_3}|$ is shown in \cref{fig:oneoverr}. We find a   transition in the expectation value  for increasing  mass ratio. For dimer states, two electrons are relatively distant, rendering the value of  $\braket{r_{B}/|\mathbf{r_2}-\mathbf{r_3}|}$ small. In contrast, for trimer states, this value is considerable and increases as the mass ratio rises. The moderate increase  of the Coulomb energy in the trimer state as function of the mass ratio,   suggests already in  this simple estimate that the existence of the dimer-to-trimer transition will persist even in presence of Coulomb repulsion.

Motivated by the above, we now solve numerically for the ground states of the system including the Coulomb interaction~\eqref{eq:Coulomb} by applying the SVM for different values of a dimensionless effective charge $q$, defined by the square root of the ratio of Coulomb repulsion to dimer binding energy 
\begin{equation}\label{eq:dimlessq}
q= \sqrt{\frac{V_{\text{FF}}(r_B)}{E_{2B}^{\infty}}} = \sqrt{\frac{2 m_{\mathrm F} r_{B}}{4 \pi \epsilon_0 \epsilon \hbar^2}} e,
\end{equation}  
where we have restored the factor of $\hbar$ for clarity.

\begin{figure}[t!]
\centering
\includegraphics[width=\linewidth]{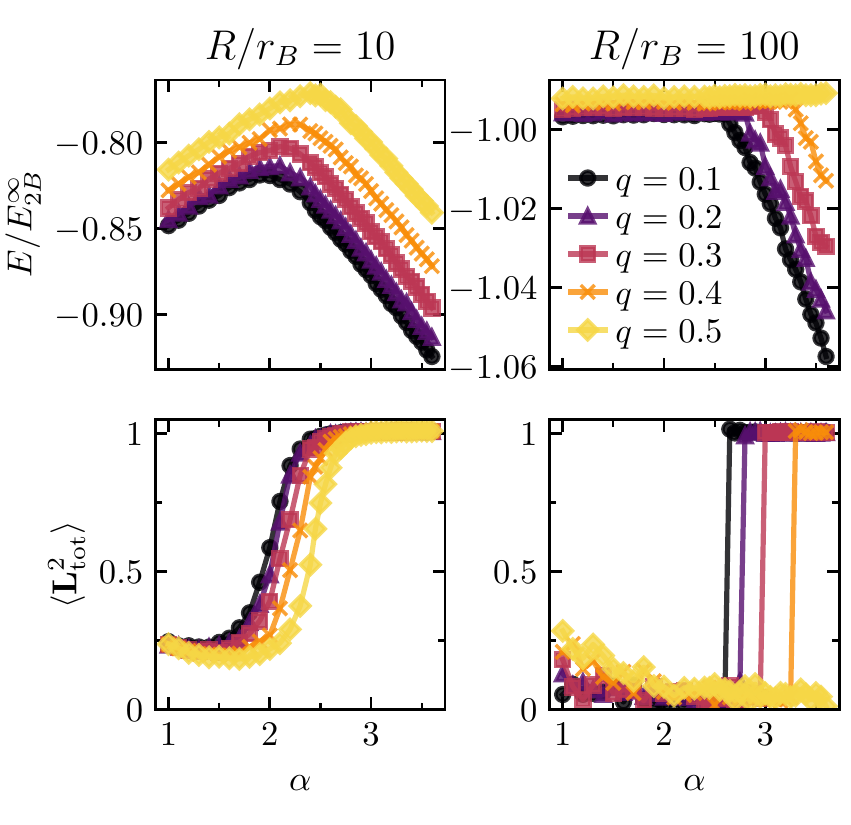}
\caption{Energies and expectation values of $\mathbf{L}^2_{\text{tot}}$ of the ground state of the system in presence of Coulomb repulsion (parametrized by the effective charge $q$) for system sizes $R/r_{B}=10$ (left), and $100$ (right). The interaction range of the fermion-impurity potential is chosen as $r_0/r_{B}=0.8$.   As in \cref{energyangular}, the steep decrease in energy beyond a critical mass ratio  reflects the crossover from a dimer  to a trimer state. The position, at which this transition occurs, moves to higher $\alpha$ upon increasing the box size $R$ and the effective charge $q$.}\label{fig:fullcoulomb}
\end{figure}

From  the SVM, we calculate the energy and the expectation value
of $\mathbf{L}^2_{\text{tot}}$ for an interaction range $r_0/r_{B}=0.8$ and box sizes $R/r_{B}=10$ and  $R/r_{B}=100$. The result is shown in  \cref{fig:fullcoulomb}. Depending on the effective charge $q$, the energies start to decrease significantly beyond a critical mass ratio. At the same time,  the corresponding values of $\mathbf{L}^2_{\text{tot}}$  rapidly increase, signaling a dimer-to-trimer crossover.

The larger the effective  charge $q$, the larger the critical value $\alpha_c$ becomes. Conversely, the larger the density $n_F$ ($\sim 1/R^2 \sim k_F^2$), the smaller the critical value of $\alpha_c$. 
Notably, the dimer-to-trimer transition remains robust upon  the strong, long-range Coulomb repulsion. Thus, while Coulomb repulsion weakens trimer formation (increasing the critical value), it does not inhibit it. Indeed, for all effective charges we considered ~\footnote{the $q=0.5$, $R/r_B=100$ data set shown in \cref{fig:fullcoulomb} does not show a trimer state, however, this is merely due to the chosen plot range. A trimer state appears eventually upon increasing the mass ratio.},  we have observed the eventual transition into a trimer state. Importantly, one can also always offset the detrimental effects of Coulomb repulsion on forming a trimer, either by tighter confinement (i.e. larger effective electron density), or a larger interaction range.

\begin{figure}[t!]
\centering
\includegraphics[width=\linewidth]{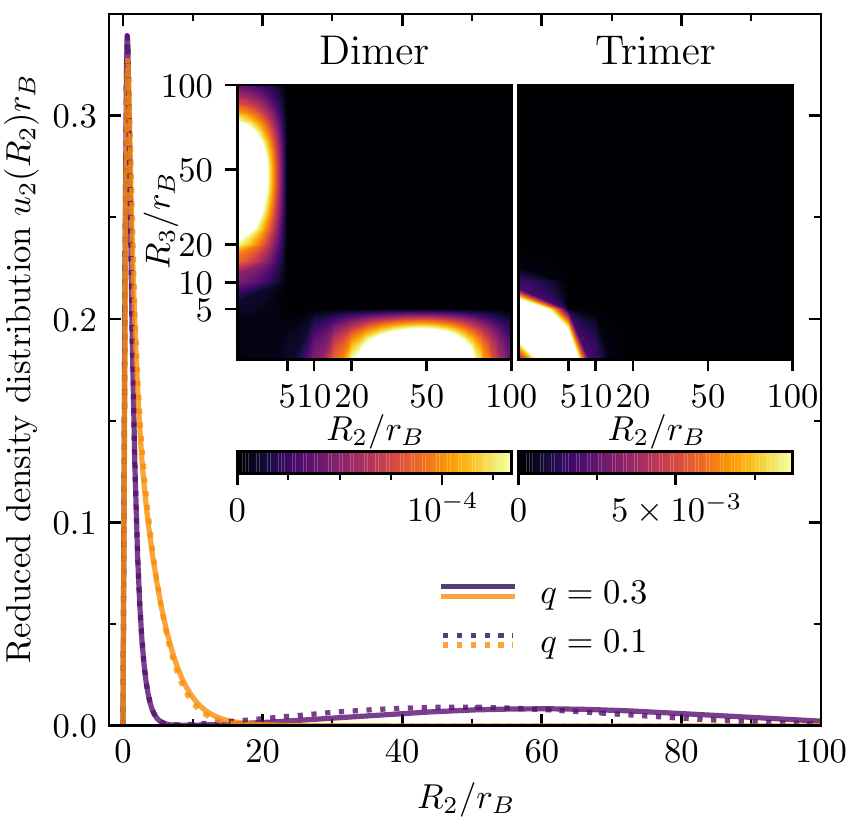}
\caption{Reduced density distributions of a dimer ($\alpha=2$) and trimer ($\alpha=3.5$) state in presence of Coulomb repulsion. The plot shows $u_2(R_2)r_B$ for a dimer (purple, solid) and a trimer (orange, solid) state for $R/r_B=100$, $r_0/r_B=0.8$ and $q=0.3$. For comparison the result is also shown for a smaller value of effective charge $q=0.1$ in orange. The  insets show $u_1(R_2,R_3)r_B^4$ for the trimer ($\alpha=3.5$, right) and the dimer state ($\alpha=2$, right) with $R/r_B=100$, $r_0/r_B=0.8$ and $q=0.3$. As in \cref{densitydist}, the qualitative distribution of fermions within the dimer and the trimer state is visible.}\label{densitydistcoulomb}
\end{figure}

We show the reduced density distribution for the  system in presence of Coulomb repulsion in \cref{densitydistcoulomb}.  The effective charges and mass ratios were chosen to realize both dimer and trimer states as in \cref{densitydist}. As  can be seen, both states feature a localized part, while the dimer again exhibits the additional contribution of a delocalized  scattering state. The density plots of $u_1$, shown in the inset of \cref{densitydistcoulomb}, exhibit the same qualitative behaviour as those in \cref{densitydist}; for a further analysis of the angular distribution of the states we refer to \cref{apptheta}. \cref{densitydistcoulomb} also shows that, increasing the effective charge $q$, the density distribution of the trimer decays over a larger length scale. This clearly shows that the Fermi-Fermi repulsion tends to favor a larger separation between fermions, while still  supporting the formation of a trimer state. Similarly, within the dimer state,  Coulomb repulsion has the effect of pushing the scattering tail away from the impurity-fermion bound state.

 For  typical parameters and energy scales in TMDs, i.e. $\epsilon \approx 4.4$, $m_{\mathrm {F}} \approx 0.5 \, m_{\mathrm {e}}$, where $m_{\mathrm {e}}$ indicates the bare electron mass,
and $|E^{\infty}_{2B}| \approx 30$ meV (trion binding energy)~\cite{Fey2020}, one arrives at $q \approx 2.6$. This value is consistent with the absence of  experimental observations of higher-order bound states  as the ground state. While at first sight this might suggest the absence of the $p$-wave trimer state for typical TMD realizations, this estimate is obtained assuming an electronic system at vanishing density. In this regard, it is important to note that, as we also find, confinement  naturally decreases the role of Coulomb interaction. In turn, regarding the increase in confinement as an increase in the effective electron density, our results suggest that at sufficiently high fermion densities,  $p$-wave bosonic trimers could indeed be stabilized as the actual ground state in the system already for the typical experimental parameters. Moreover, our results show that the critical mass ratio $\alpha_c$ could be  changed by experimentally tuning the effective charge $q$. This could, for instance, be realized by appropriate dielectric engineering of the materials \cite{Steinleitner18} that encapsulate the TMD layer \footnote{Such a modification of the dielectric environment  will also affect the trion binding energy resulting  in  a redefinition of $E_{2B}^\infty$.}.

\section{Discussion and outlook}

We have studied the influence of confinement and finite interaction ranges on the formation of ground state trimers in confined three-body systems where two identical fermions interact with a mobile quantum impurity. We have shown that the position of the dimer-to-trimer transition, previously characterized in Refs.~\cite{Pricoupenko2010,Jesper2013,Cui2022}, varies significantly under these effects. Our results show how these effects can, in principle, be leveraged to realize $p$-wave trimers in atomically-thin semiconductors and ultracold quantum gases. 

While in two-dimensional cold atom systems already a great variety of mass ratios is available, trimer formation could be further enhanced using a longitudinal trapping confinement. In TMDs,  the available mass ratios are more restricted (unless, e.g., flat  Moiré bands are considered). However, our results show that a finite exciton-electron interaction range as well as  confinement can enhance and stabilize trimer formation. Furthermore, we have argued that, given a suitable TMD,  trimers can, in principle, survive Coulomb repulsion as long as the effective charge, given by material parameters such as the dielectric constant, remains below a critical value.

In regards to interpreting confinement as a means to imitate a finite bath density, the remarkable robustness of the dimer-to-trimer transition suggests that bosonic $p$-wave trimers might already appear as the ground state of realistic TMD heterostructures
\cite{wang2018tmdspectroscopy}. Our work thus highlights that experiments may already  be close to the point of exploring exciton-induced  $p$-wave electron pairing, opening up the avenue to  novel mechanisms of exciton-mediated $p$-wave superconductivity in van-der Waals materials.  

Moving forward from our work, there are  several further exciting paths to pursue. For one, it has been shown that for systems with a greater number of bath particles also higher-order bound states may play an important role \cite{Cui2022}, which could lie lower in energy than the trimer state. The influence  of confinement and finite range on these states is unexplored, and might drastically change the position of ground state transitions as well as the occurrence of these transitions in the first place. 

Going beyond $1+N$-type systems, the phase diagram of Bose-Fermi  mixtures \cite{duda2023} at a given density imbalance of the constituent species might be studied in few-body systems with comparable density ratios. In this regard, the occurrence, nature, and dynamics of interesting phenomena such as phase separation in the many-body regime could be illuminated by corresponding observations in a few-body system. For instance, in a system of type $2+3$, one might compare the formation of a four- or five-particle bound state to the coexistence of a dimer with a trimer. 

Cold atomic systems offer a wealth of tunable parameters such as  mass ratio, bound-state energy and confinement \cite{bloch2008}. Moreover, ultracold polar molecules and magnetic atoms with strong dipolar interactions can now be realized experimentally~\cite{Moses2016,Bohn2017,Chomaz2022}. Exploiting the long-range character of their interactions, the effects of Coulomb repulsion between identical fermions in solid-state structures can now be mimicked in cold atom systems, highlighting these as an exciting platform to gain new insights into the physics of the exciton-electron mixtures in layered van der Waals materials.

\section{Acknowledgements}
We thank Selim Jochim, Christian Fey and Mikhail Glazov for inspiring discussions. We acknowledge support by the Deutsche Forschungsgemeinschaft under Germany's Excellence Strategy EXC 2181/1 - 390900948 (the Heidelberg STRUCTURES Excellence Cluster). The work at ETH Zurich was supported by the Swiss National Science Foundation (SNSF) under Grant Number 200021-204076.
  J.~v.~M. is supported by a fellowship of the International Max Planck Research School for Quantum Science and Technology (IMPRS-QST).


%

\appendix

\section{DETAILED DESCRIPTION OF THE SVM ALGORITHM}\label{svmalgorithm}

\subsection{Algorithm and Sampling}

In this appendix, we provide further information on the optimization process undertaken in every step of the SVM \cite{suzuki1998}. For the results shown in the main text, we perform 10 independent calculations for every data point. In each of these calculations, 100 basis states are
computed. In the following, we refer to each one of these calculations as a run, and the combination of 10 runs makes up a single data point. 

To compile a set of 100 basis states $\left \{ \Phi_n \right\}_{n=1}^{100}$   in a single run, we successively increase the set of basis states by drawing from the manifold of trial wavefunctions described in the main text. In a step $N\to N+1$, we draw proposal states $\left \{\Phi_\alpha\right\}$ independently. From these proposals, we choose the state  $\Phi_\beta$ which produces the lowest-lying eigenstate of the Hamiltonian $H$ with respect to the vector space $V_\alpha^N$ spanned by the states $\left \{ \Phi_n \right\}_{n=1}^{N} \cup \Phi_\alpha$ (for a detailed description of SVM and different optimization strategies see Ref. \cite{suzuki1998}). Specifically,
\begin{align}
V_\alpha^N &=  \spn{\left(\left \{ \Phi_n \right\}_{n=1}^{N} \cup \Phi_\alpha\right)}   \\
\{\lambda_{\alpha,1}^N,...,\lambda_{\alpha,N}^N \}&= \sigma(H|_{V_\alpha^N}))\\
\beta&= \min_{\alpha}{\left[ \min_i(\{\lambda_{\alpha,i}^N\}_i)\right]},
\end{align}
where $\sigma(H|_{V_\alpha^N}))$ denotes the spectrum of the Hamiltonian $H$, restricted to the vector space $V_\alpha^N$. The minimization over $i$ chooses the lowest eigenvalue of $H|_{V_\alpha^N}$, while the minimization over $\alpha$ optimizes the proposal state.

Next, we perform a random descent walk in the vicinity of $\Phi_\beta$, for which every step is accepted so long as it lowers the lowest eigenvalue. This process is terminated after a fixed number of proposals (specified below).

A straightforward method to draw independently from the ECG  manifold is to draw proposal states $\Phi_\alpha$ as 
\begin{align}
m_\alpha= \frac{1}{ R} \left(\begin{matrix} x_{11}& x_{12}& x_{13}\\ 
 x_{21}& x_{22}& x_{23}\\ 
  x_{31}& x_{32}& x_{33}
\end{matrix}\right)\label{dimersampling}
\end{align} 
with 
\begin{align}
A_\alpha= m_\alpha^T m_\alpha^{\phantom{T}}.
\end{align}
Here, the $x_{ij}$ are drawn from a uniform distribution in the interval $x_{ij}\in[-1,1]$. The corresponding (unrenormalized) basis state is then given as $\Phi_\alpha(\mathbf{r}_1,\mathbf{r}_2,\mathbf{r}_3)r_B^3=\mathcal{P}\exp{\left(-\frac{1}{2}\sum_{i,j=1}^3A_{\alpha,ij}\mathbf{r}_i\cdot\mathbf{r}_j\right)}$. In the second part of the optimization, in which we perform the random descent walk,  the proposal is updated as 
\begin{align}
m'_{\beta}= m^{\phantom{'}}_\beta + \delta x \left(\begin{matrix} x_{11}& x_{12}& x_{13}\\ 
 x_{21}& x_{22}& x_{23}\\ 
  x_{31}& x_{32}& x_{33}
\end{matrix}\right)\label{othersampling}
\end{align}
with 
\begin{align}
A'_\beta=(m_\beta')^T m_\beta'.
\end{align} 
In practice,  a value of $\delta x=0.1/r_B$ has shown to yield good results for the parameters considered in this work.

As the manifold of trial functions is fairly large, a large number of random proposals is necessary in every step of the algorithm to ensure convergence. This choice of sampling quickly yields reliable results for dimer states. For trimer states, convergence is much slower, and especially close to the dimer-to-trimer transition, it can occur that no trimer state is obtained. To address this challenge,  we leverage the physical intuition that a trimer state should feature all  particles confined within a length scale of the interaction range. Exploiting this fact also allows us to  reduce the number of required steps, as well as improve stability of the algorithm. 

In order to  implement this idea in the algorithm, it is important to note that the matrices $A^{-1}_\alpha$ carry the meaning of a covariance matrix. This  suggests a covariance matrix of close to constant value (proportional to the mean distance squared of the particles from the center of the trap), with fluctuations around this value of the order of the interaction range. We thus introduce a further sampling method described as 
\begin{align}
B&= 5 R x_0+ 2 r_0 \left(\begin{matrix} x_{11}& x_{12}& x_{13}\\ 
 x_{21}& x_{22}& x_{23}\\ 
  x_{31}& x_{32}& x_{33}
\end{matrix}\right)\nonumber , \\
A_\alpha^{-1}&= \frac{B+ B^T}{2} , \label{trimersamplingnew}
\end{align}
where $x_0\in [0, 1] r_B$, $x_{i,j}\in [-1,1] r_B$.  Using $\delta x'=0.1 r_B$ the corresponding random walk method is determined by 
\begin{align}
B'&= B+ \delta x' \left(\begin{matrix} x_{11}& x_{12}& x_{13}\\ 
 x_{21}& x_{22}& x_{23}\\ 
  x_{31}& x_{32}& x_{33}\end{matrix}\right)\nonumber , \\
(A'_\alpha)^{-1}&= \frac{B'+ (B')^T}{2} .\label{trimersampling}
\end{align}
In \cref{trimersamplingnew}, the value of $5R$ (representing  a sampling range of the mean distance squared of the particle from the center of the trap) was chosen to ensure reasonable convergence. The value of $2r_0$, in turn, representing the interparticle distances was selected based on the fact that the localization of particles with respect to  each other should be on the order of the interaction range.
In the first part of the algorithm, where states are independently drawn, we then alternate between the two sampling methods, \cref{dimersampling,trimersamplingnew}, while in the second part (where the random walk is performed) we alternate between the methods defined by  \cref{othersampling,trimersampling}.

While the  sampling method described by \cref{trimersamplingnew} may seem very biased at first glance, this presumption does not capture the full picture for several reasons. First, states of the form of \cref{trimersamplingnew}  are usually also found using the sampling method described in \cref{dimersampling}. There, however, many more sampling steps are necessary for this to occur, implying less efficiency. Second, states found using \cref{trimersamplingnew} are often accepted in the regime where the trimer is the ground state, but indeed also where the dimer is the ground state. 
Furthermore, we use a large number of sampling steps, namely about 15000 independent samples and 15000 local descents, each repeated twenty times for every run.  Thus the exact form of the sampling coefficients used in \cref{dimersampling,trimersampling} does not play a  dominant role as long as the space of eligible wavefunctions is sufficiently small (\emph{i.e.} the confinement is not too big) and the space of  appropriate wavefunction is large enough (\emph{i.e.} the interaction range is not too small). However, as can be seen in \cref{energyangular}, for some parameter regimes, the data begins to develop a scatter which could be addressed by increasing the number of sampling steps further, or by restricting the sampling method described in \cref{trimersampling} to a smaller parameter space.

After we have finally performed 10 different runs, each yielding 100 basis states, we then combine the results of these different runs to obtain a basis set of 1000 basis states $\left \{ \Phi_n \right\}_{n=1}^{1000}$. In the very end, the Hamiltonian is diagonalized with respect to these 1000 states and the physical quantities are extracted from the resulting ground state. These results are shown  in the main text.

\subsection{The Hamiltonian in the ECG basis}
Here we give a detailed account of the representation of the Hamiltonian in \cref{hamiltonian} within the manifold spanned by the ECG. 

Given two basis functions $|A\rangle$ and $|B\rangle$, corresponding to  $\langle \mathbf{x}|A\rangle r_B^3=\exp{\left(-\frac{1}{2}\sum_{i,j=1}^3A_{ij}\mathbf{r}_i\cdot\mathbf{r}_j\right)}$ and $\langle \mathbf{x}|B\rangle r_B^3=\exp{\left(-\frac{1}{2}\sum_{i,j=1}^3B_{ij}\mathbf{r}_i\cdot\mathbf{r}_j\right)}$ with $\mathbf{x}^{\text{T}}=(\mathbf{r}_1^{\text{T}},\mathbf{r}_2^{\text{T}},\mathbf{r}_3^{\text{T}})$, the matrix elements of the kinetic parts  $-\frac{\hbar^2}{2m_i}\nabla_i^2=\frac{\mathbf{p}_i^2}{2m_i}$ are given by \cite{Varga2008}
\begin{equation}
\langle A|\mathbf{p}_i^2|B\rangle r_B^6=\frac{16\pi^3}{\det{(A+B)}}[A(A+B)^{-1}B]_{ii}.
\end{equation}

The remaining parts of the Hamiltonian consist of one-and two-body potentials of the form $V(\mathbf{r}_i)$ and $V(\mathbf{r}_i-\mathbf{r}_j)$. Given a suitable vector $w^{\text{T}}=(w_1,w_2,w_3)$,  both types of potential can thus be written in the form $V(\tilde{w}^{\text{T}}\mathbf{x})$ where $\tilde{w}^{\text{T}}=w^{\text{T}}\otimes\mathbb{I}_{2\times2}$ such that $\tilde{w}^{\text{T}}\mathbf{x}= w_1 \mathbf{r}_1+  w_2 \mathbf{r}_2+  w_3 \mathbf{r}_3$. 
Then, the matrix element of this general form of the potential reads  \cite{Varga2008,Fedorov2016}
\begin{equation}
\begin{aligned}
\langle A|V(\tilde{w}^{\text{T}}\mathbf{x})|B\rangle=&\frac{4\pi^2}{\det{(A+B)}}\frac{r_B^{-6}}{w^{\text{T}}(A+B)^{-1}w}\\
&\times\int d\mathbf{r}V(\mathbf{r})\exp{\left(-\frac{1}{2w^{\text{T}}(A+B)^{-1}w}\mathbf{r}^2\right)} .
\end{aligned}
\end{equation}

From this expression, the matrix elements of $V_{\text{conf}}(\mathbf{r}_i)=E_{2B}^{\infty}(|\mathbf{r}_i|/R)^n$, $V_{\text{FI}}(\mathbf{r}_i-\mathbf{r}_j)$, and $V_{\text{FF}}(\mathbf{r}_i-\mathbf{r}_j)$ can be obtained for appropriate choices of $w$. For $V_{\text{conf}}(\mathbf{r}_i)$, 
 $w_j=\delta_{ij}$, while for $V_{\text{FI}}(\mathbf{r}_i-\mathbf{r}_j)$ and $V_{\text{FF}}(\mathbf{r}_i-\mathbf{r}_j)$, $w_k=\delta_{ik}-\delta_{jk}$. 

The matrix element of $V_{\text{conf}}(\mathbf{r}_i)$ is then given by 
\begin{align}
    \langle A|V_{\text{conf}}(\mathbf{r}_i)|B\rangle
    &=\frac{E_{2B}^{\infty}}{r_B^6}\frac{8\pi^3}{\det{(A+B)}}\frac{a_i^{-n/2}}{R^n}\Gamma{\left(\frac{n+2}{2}\right)} \  ,
\end{align}
where  $a_i=1/2(A+B)^{-1}_{ii}$ and  $\Gamma{(x)}$ is the Gamma function. In our calculation we use $n=30$.

The matrix element of the fermion-impurity interaction reads 
\begin{equation}
\begin{aligned}
    \langle A|V_{\text{FI}}(\mathbf{r}_i-\mathbf{r}_j)|B\rangle
    &=-\frac{8\pi^3}{r_B^6 \det{(A+B)}}V_0\left(1-e^{-b_{ij}r_0^2}\right),
\end{aligned}
\end{equation}
 where $b_{ij}=1/2w^{\text{T}}(A+B)^{-1}w=1/2((A+B)^{-1}_{ii}+(A+B)^{-1}_{jj}-2(A+B)^{-1}_{ij})$. 
 
 Finally, the matrix element of $V_{\text{FF}}(\mathbf{r})=E_{2B}^\infty r_B q^2 /|\mathbf{r}|$ is given by

\begin{align}
    \langle A|V_{\text{FF}}(\mathbf{r}_2-\mathbf{r}_3)|B\rangle r_B^6
    &=c\frac{8\pi^3}{\det{(A+B)}}\sqrt{\pi b_{23}} ,
\end{align}

where $c= E_{2B}^\infty r_B q^2$. 

\subsection{Angular Momentum in the ECG basis}
	The total angular momentum of the (2+1) system relative to the impurity is given by $\mathbf{L}_{\text{tot}}=\mathbf{L}_2+\mathbf{L}_3=\mathbf{R}_2\times\mathbf{P}_2+\mathbf{R}_3\times\mathbf{P}_3$, where $\mathbf{R}_2$, $\mathbf{R}_3$, $\mathbf{P}_2$, $\mathbf{P}_3$ are the positions and momenta of the two fermions relative to the impurity. Because our variational wavefunctions are  real functions, it follows that $\langle \mathbf{L}_{\text{tot}}\rangle=0$~\cite{suzuki1998} . 
    
    In order to capture the transition from the dimer state (with $\langle\mathbf{L}_{\text{tot}}\rangle=0$) to the trimer state (with $\langle\mathbf{L}_{\text{tot}}\rangle=\pm1$), we focus on the expectation value of $\mathbf{L}_{\text{tot}}^2$. To that end, we first define the coordinate transformation  $\mathbf{R}_1=m_I\mathbf{r}_1/(m_I+2m_F)+m_F\mathbf{r}_2/(m_I+2m_F)+m_F\mathbf{r}_3/(m_I+2m_F)$, $\mathbf{R}_2=\mathbf{r}_2-\mathbf{r}_1$, $\mathbf{R}_3=\mathbf{r}_3-\mathbf{r}_1$. Given an ECG wavefunction $|A\rangle$ with $\langle \mathbf{x}|A\rangle r_B^3=\exp{\left(-\frac{1}{2}\sum_{i,j=1}^3A_{ij}\mathbf{r}_i\cdot\mathbf{r}_j\right)}$ and $\mathbf{x}^{\text{T}}=(\mathbf{r}_1^{\text{T}},\mathbf{r}_2^{\text{T}},\mathbf{r}_3^{\text{T}})$, this can be represented in the relative coordinates as $\langle\tilde{\mathbf{x}}|A\rangle r_B^3=\exp{\left(-\frac{1}{2}\sum_{i,j=1}^3\tilde{A}_{ij}\mathbf{R}_i\cdot\mathbf{R}_j\right)}$ where $\tilde{\mathbf{x}}^{\text{T}}=(\mathbf{R}_1^{\text{T}},\mathbf{R}_2^{\text{T}},\mathbf{R}_3^{\text{T}})$, and $\tilde{A}$ is defined as
	\begin{equation}
	    \tilde{A}=O^{\text{T}}AO
	\end{equation}
 with
	\begin{equation}
	O=\left(\begin{matrix}
	\frac{m_I}{m_I+2m_F} & \frac{m_F}{m_I+2m_F} & \frac{m_F}{m_I+2m_F} \\
	-1 & 1 & 0 \\
	-1 & 0 & 1\\
	\end{matrix}
	\right)^{-1}.
    \end{equation}
	The matrix element of $\mathbf{L}_{\text{tot}}^2$ is then given as (for more detail see Ref. \cite{suzuki1998})
	
	\begin{widetext}
	\begin{equation}\label{eq:lsquare}
	\begin{aligned}
	\langle A|\mathbf{L}_{\text{tot}}^2|B\rangle r_B^6
    &=\frac{8\pi^3}{\det{(\tilde{A}+\tilde{B})}}\left(\frac{1}{4}\text{Tr}\left(f(\tilde{A})((\tilde{A}+\tilde{B})^{-1}\otimes\mathbb{I}_{2\times2})\right)\text{Tr}\left(f(\tilde{B})((\tilde{A}+\tilde{B})^{-1}\otimes\mathbb{I}_{2\times2})\right)\right.\\
    &\left.+\frac{1}{2}\text{Tr}\left(f(\tilde{A})((\tilde{A}+\tilde{B})^{-1}\otimes\mathbb{I}_{2\times2})f(\tilde{B})((\tilde{A}+\tilde{B})^{-1}\otimes\mathbb{I}_{2\times2})\right)\right).
	\end{aligned}
	\end{equation}
	\end{widetext}
	Here, we have defined the function $f(\tilde{A})$ in the following way. Given a symmetric $3\times3$  matrix 
	\begin{equation}
	\tilde{A}=\left(\begin{matrix}
	\tilde{A}_{11} & \tilde{A}_{12} & \tilde{A}_{13} \\
	\tilde{A}_{12} & \tilde{A}_{22} & \tilde{A}_{23} \\
	\tilde{A}_{13} & \tilde{A}_{23} & \tilde{A}_{33}\\
	\end{matrix}
	\right),
	\end{equation} 
    we define  $R=\left(\begin{matrix}
	0 & 1 \\
	-1 & 0 \\
	\end{matrix}
	\right)$, such that  $f(\tilde{A})$ reads
	\begin{equation}
	f(\tilde{A})=\left(\begin{matrix}
	0 & -\tilde{A}_{12} & -\tilde{A}_{13} \\
	\tilde{A}_{12} & 0 & 0 \\
	\tilde{A}_{13} & 0 & 0\\
	\end{matrix}
	\right)\otimes R \ .
	\end{equation}

\section{TWO-BODY PROBLEM}\label{twobodyall}
In this appendix, we show the text-book solution of the two-body problem in absence of external confinement, and then go on to study the influence of confinement on the solution of the two-body problem.
\subsection{Without confinement}
\label{twobody}
We consider an impurity (with mass $m_I$) interacting with a single fermion (with mass $m_F$) via $V_{\text{FI}}(\mathbf{r})=-V_0\theta(r_0-|\mathbf{r}|)$ where $\theta(x)$ is the Heaviside function. The Schroedinger equation in the relative coordinate frame reads
\begin{equation}
    -\frac{\nabla^2}{2\mu}\psi(\mathbf{r})+V_{\text{FI}}(\mathbf{r})\psi(\mathbf{r})=-E^{\infty}_{2B}\psi(\mathbf{r})
\end{equation}
where $\mu=m_Fm_I/(m_F+m_I)$ is the reduced mass of an impurity and a fermion. The wavefunction $\psi(\mathbf{r})$ can be decomposed into a radial part and an angular part, i.e. $\psi(\mathbf{r})=u(r)e^{im\theta}$ with $m$ the angular momentum of the state. For the ground state, we have $m=0$. Thus the equation for the radial wavefunction is given by
\begin{equation}
    r^2u''+ru'+2\mu[-E^{\infty}_{2B}+V_0\theta(r_0-r)]r^2u=0.
\end{equation}
The solution of this equation can be found in text books \cite{Whitehead2016}. The ground state energy $E^{\infty}_{2B}$ is found by the solution of the implicit equation 
\begin{widetext}
\begin{equation}
\begin{aligned}
  \sqrt{2\mu E^{\infty}_{2B}}J_0\left(\sqrt{2\mu(-E^{\infty}_{2B}+V_0)}r_0\right)K_1\left(\sqrt{2\mu E^{\infty}_{2B}}r_0\right)
    -\sqrt{2\mu(-E^{\infty}_{2B}+V_0)}K_0\left(\sqrt{2\mu E^{\infty}_{2B}}r_0\right)&J_1\left(\sqrt{2\mu(-E^{\infty}_{2B}+V_0)}r_0\right)\\
    &=0,
\end{aligned}
\end{equation}
\end{widetext}
with $J_0$, $J_1$ Bessel functions of the first kind, and $K_0$, $K_1$ modified Bessel functions of the second kind.

\subsection{With confinement}
\label{apptwobodyconf}
\begin{figure}[t!]
\centering
\includegraphics[width=\linewidth]{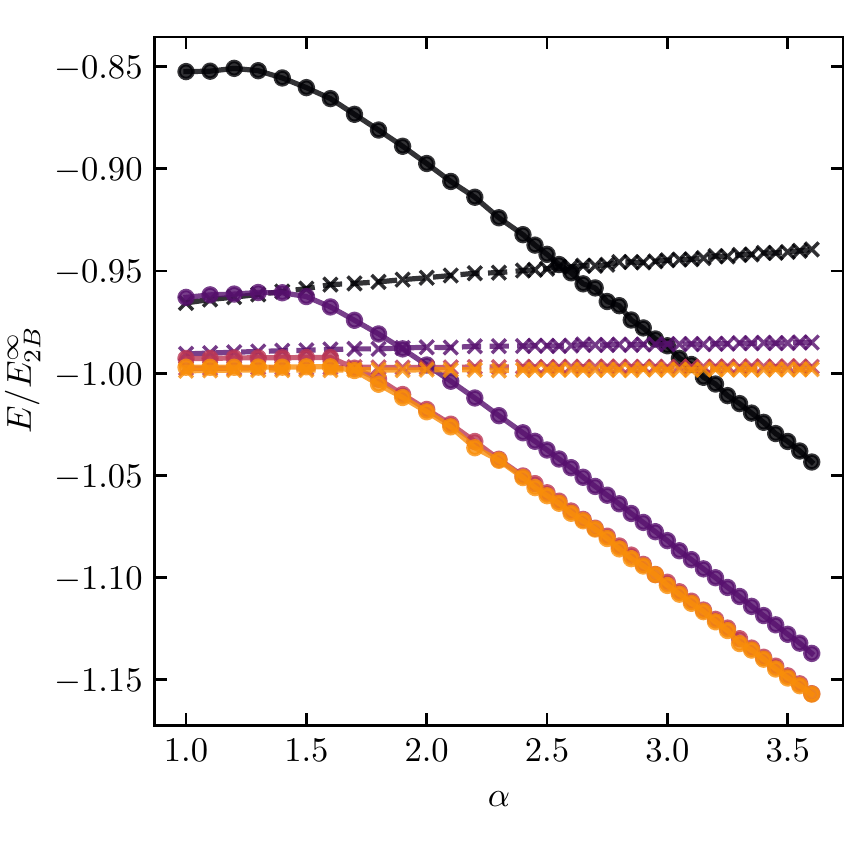}
\caption{Ground state energy of the two- (crosses, dashed lines) and three-body system (dots, solid lines) for  $R/r_B=10$ (black), $20$ (purple), $50$ (red) and $100$ (orange), and a fixed value $r_0/r_B=1.2$. The two-body ground state energy increases linearly with mass ratio $\alpha$, which coincides with the behavior of the three-body ground state energy below the critical mass ratio.}\label{twobodyenergy}
\end{figure}
We next consider a two-body system consisting of one impurity (with mass $m_I$) and one fermion (with mass $m_F$) in a 2D spherical box. The Hamiltonian then reads
\begin{equation}
H=-\frac{\hbar^2}{2m_I}\nabla_1^2-\frac{\hbar^2}{2m_F}\nabla_2^2 +\sum_{i=1}^2 V_{\text{conf}} (\textbf{r}_i) 
+V_{\text{FI}}(\textbf{r}_1-\textbf{r}_2),
\end{equation}
where we have used the same notation as in \cref{twobody}. To solve this two-body problem, we employ the SVM as described in the main text.

In \cref{twobodyenergy}, we show the two-body  as well as the three-body ground state energy as a function of $\alpha$ for $r_0/r_B=1.2$ and different values of $R$. The  dimer energies lie slightly higher than $-E_{2B}^{\infty}$ due to the confinement, while for larger system sizes the energies approach $-E_{2B}^{\infty}$. Additionally, a close to linear increase of the energies with the mass ratio $\alpha$ is visible, which decreases as $R$ increases. This observation is  in line with the interpretation of a decrease of the two-body confinement energy, given by $E_{\text{conf}}=z_{01}^2/2m_IR^2+z_{01}^2/2m_FR^2=z_{01}^2(\alpha+1)/2m_FR^2$, where $z_{01}$ is the first zero of the Bessel function $J_0$. 

Comparing the three-body energy with the two-body energy, one can see that,  below the critical mass ratio, the three-body energy also increases linearly with $\alpha$, and that the increase is larger for smaller box size. Additionally, especially for smaller system sizes, the three-body energies below the critical mass ratio lie considerably higher than their two-body counterparts. This is caused by the confinement energy of the fermion in a scattering state, as expected from our analysis in  \cref{nonintferm}.

\section{CONVERGENCE ANALYSIS}\label{app:convergence_analysis}
In this appendix, we analyze the deviation of the results shown in \cref{energyangular} from the asymptotic critical mass ratio $\alpha_{c}\approx3.34$ \cite{Pricoupenko2010,Jesper2013, Cui2022} obtained for $R\to \infty$ and $r_0\to 0$. We then analyze the convergence of the results shown in \cref{energyangular}, which is followed by a study with regards to the number of wave functions sampled in every expansion step. 

\subsection{Deviation from the asymptotic result $\alpha_{c}\approx3.34$}\label{app:convergence_analysis_a}

\begin{figure}[t!]
\centering
\includegraphics[width=\linewidth]{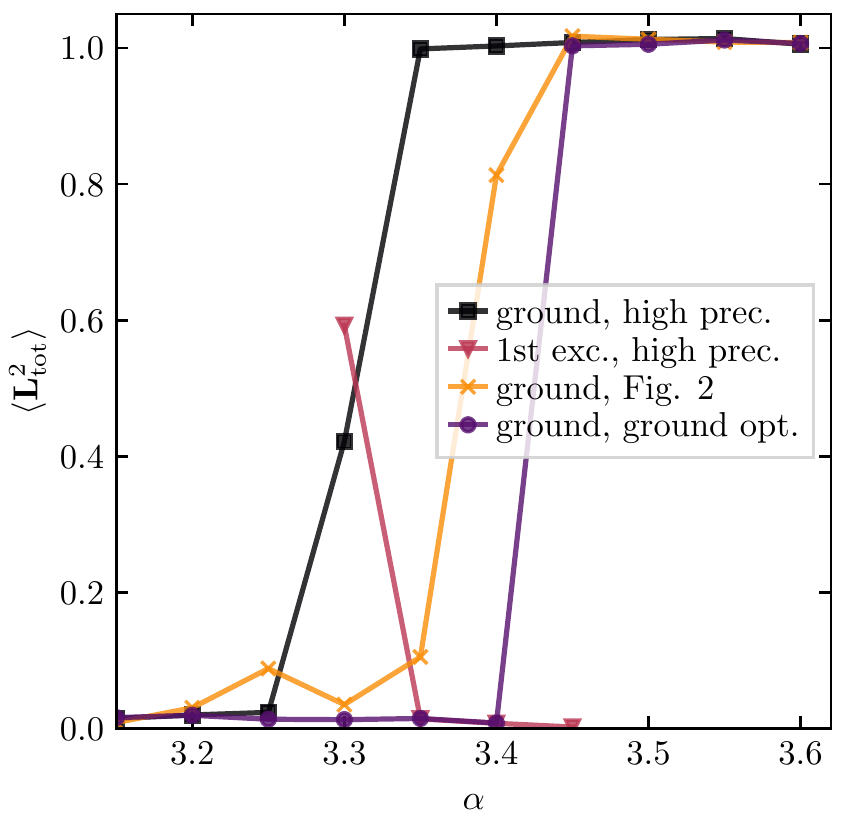}
\caption{Angular momentum values as  function of the mass imbalance $\alpha$, for  $R/r_B=100$ and $r_0/r_B=0.2$. The data of the ground state from \cref{energyangular} is shown (yellow, crosses) along with ground state data obtained from single runs with $N=1000$ basis states. This data was obtained using two different optimization strategies:  either by optimizing with respect to the ground state  (purple dots), or by optimizing for the first excited state in the second half of the run (black squares). For the latter, the corresponding expectation value  of the first excited state properties is shown as red triangles.}
\label{fig:L_ext}
\end{figure}

As it can be seen in  \cref{energyangular},  for $R/r_B=100$ and $r_0/r_B=0.2$  the $\braket{\mathbf{L}^2_{\text{tot}}}$ value begins to increase at around $\alpha=3.35$ and has arrived at approximately $1$ at the data point corresponding to $\alpha=3.45$. Thus, the data indicates that the transition occurs for $3.35<\alpha_c<3.45$, which lies higher than the asymptotic value
of $\alpha_{c}\approx3.34$. This is in opposition to our finding that, generally, confinement and a finite interaction range should in fact cause a reduction of the value of $\alpha_c$.

To study this deviation, for each  mass ratio we perform single runs of up to 1000 basis states, rather than performing ten runs of up to 100 basis states. These runs are executed in two different ways which are motivated by noting the important point  that for $R/r_B=100$, $r_0/r_B=0.2$  the transition region is very narrow (for further information, see also the detailed discussion in \cref{appendix_conv_MC}).
Hence, few basis states share overlaps with both the dimer and the trimer state. As a consequence, the ground and the excited state each have to be optimized for with a significant number of basis states, as few basis states optimize the energy of both the trimer and the dimer, and it is thus easy to miss the true ground state. 

This is visible in \cref{fig:L_ext}, where the purple dots show the result of a single run in which the expansion of the basis set towards 1000  states keeps optimizing with respect to the current ground state (and convergence is thus slow when the dimer and trimer state are almost degenerate in energy). In contrast, convergence can be dramatically sped up by allowing for more drastic updates; specifically, by adapting the acceptance criteria for basis states such that for the first 500 basis states acceptance depends on improving the ground state, and, for the next 500 basis states, it depends on improving the first excited state.  Away from the transition this is not an efficient method to obtain a good ground state estimate. However, close to the transition this approach offers dramatically improved efficiency in describing the ground state. The result is shown as black squares in \cref{fig:L_ext}. For both optimization criteria one can see that, compared to the data shown in \cref{energyangular} (reproduced also in \cref{fig:L_ext}), the scatter in $\braket{\mathbf{L}^2_{\text{tot}}}$ is absent, and the transition region has become sharper. While for the pure ground state optimization, the transition still occurs for $3.4<\alpha_c<3.45$, for the first excited state optimization criterion in the second half of the run, it now sets on shortly before $\alpha=3.3$ and  $\braket{\mathbf{L}^2_{\text{tot}}}\approx 1$ is reached shortly before $\alpha=3.35$, consistent with the free space result.

\begin{figure}[t!]
\centering
\includegraphics[width=\linewidth]{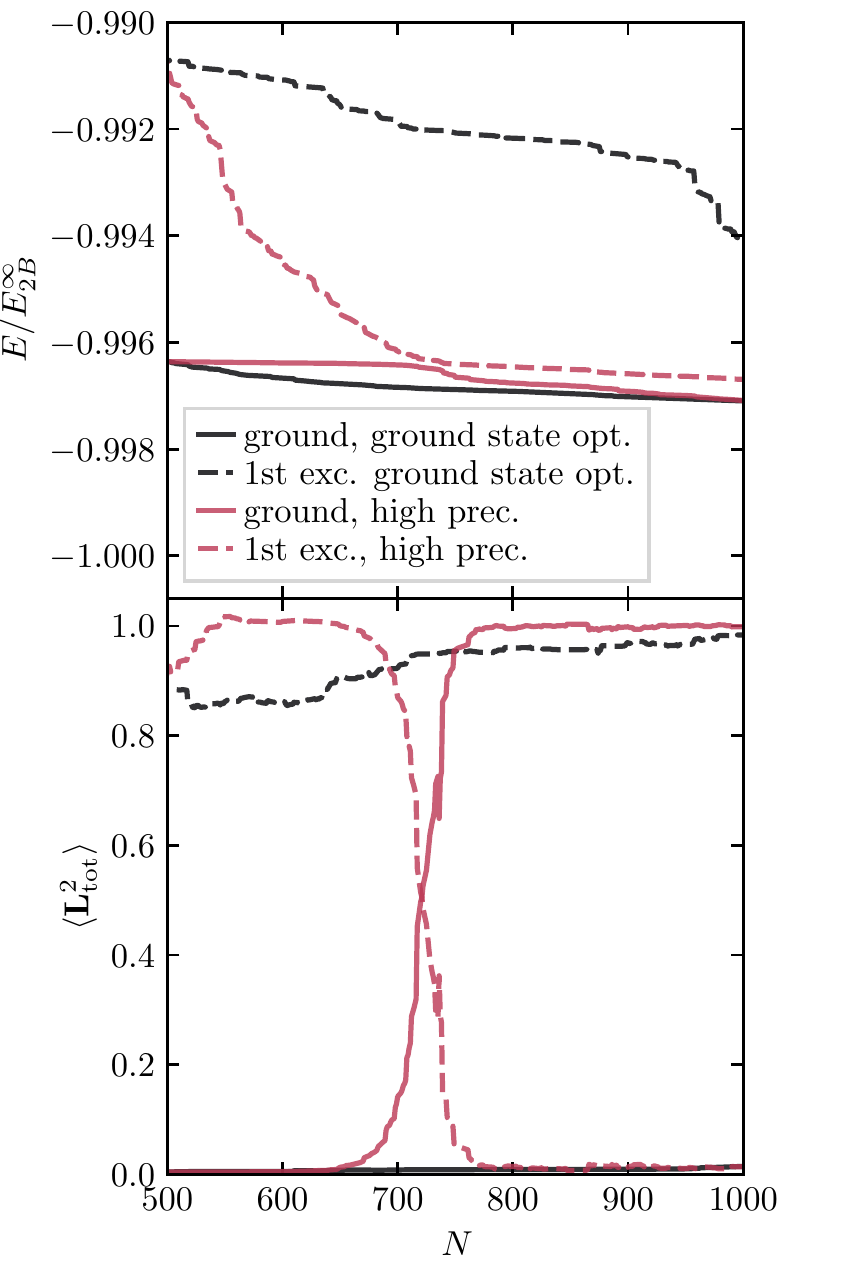}
\caption{Energy (top) and angular momentum (bottom) values of the ground and first excited state as a function of basis states $N$ for SVM calculations at $\alpha=3.35$, $R/r_B=100$ and $r_0/r_B=0.2$ using a single run of up to $1000$ basis states. The results were obtained in two different ways: optimizing the ground state for all 1000 states (black) and optimizing the ground state for 500 states and then optimizing the first excited state for another 500 states (red), showing ground (solid) and first excited state properties (dashed).}
\label{fig:verlauf}
\end{figure}
 
In order to offer further insight into the two different optimization criteria, in \cref{fig:verlauf}, energy and angular momentum data of the ground and first excited state are shown for parameters $\alpha=3.35$, $R/r_B=100$, $r_0/r_B=0.2$  close to the free-space dimer-to-trimer transition. The data, shown as a function of the number of basis states, is  obtained in the two different ways described above. That is, optimizing the ground state for all 1000 basis states (``ground state opt."), and optimizing the ground state for the first 500 basis states followed by the optimization of the first excited state for the next 500 basis states (``high prec."). By construction, the latter algorithm is  more efficient in allowing admixtures of the excited state manifold to the optimized basis set. 

As one can see from \cref{fig:verlauf}, in the first approach that optimizes for the ground state only, the energy of the ground state saturates already early on, and the first excited state sees very little improvement. Optimizing the first excited state as well, however, the energies cross over, triggering a transition from dimer to trimer behavior as can be seen in the corresponding angular momentum plot in \cref{fig:verlauf}. Here, it can also be seen that optimizing the ground state only,  its angular momentum remains close to $0$, while the first excited state does not immediately attain a value close to $1$; which is natural, since it is not optimized for. Optimizing for the first excited state in the second half of the algorithm, one can see that its expectation value attains a value close to $1$ already after being optimized for only about
100 basis states. At around 700 basis states, the first excited state has been optimized enough to trigger the crossover between ground and first excited state. 

\subsection{Convergence analysis for data shown in \cref{energyangular}}
\label{appendix_conv_MC}

 \begin{figure*}[t!]
\centering
\includegraphics[width=\linewidth]{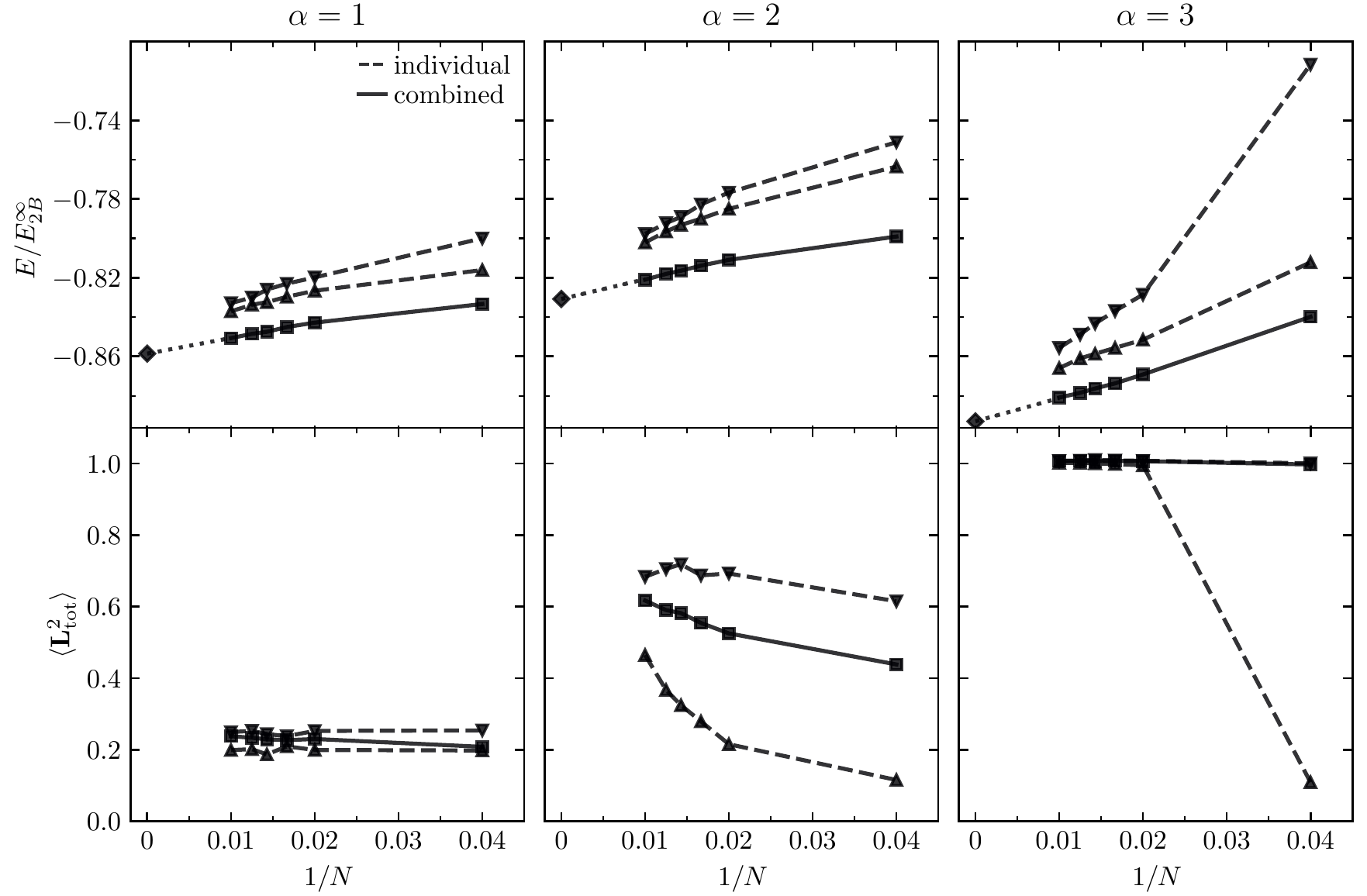}
\caption{Analysis of convergence with increasing number of basis states $N$, for parameters $R/r_B=10$, $r_0/r_B=0.8$
that yield a critical mass ratio  of around $\alpha_c=2$. The analysis is conducted for mass ratios well before (left, $\alpha=1$), close to (center, $\alpha=2$) and well after (right, $\alpha=3$) the dimer-to-trimer transition. For every data point ($\alpha$, $R/r_B$, $r_0/r_B$), shown in \cref{energyangular}, 10 independent runs with up to 100 basis states were conducted. For the present figure, the ground state in each of these 10 independent runs was tracked as the number of basis states $N$ increased from 1 to 100, and the energies and angular momentum expectation values $\langle\mathbf{L}^2_{\text{tot}}\rangle$ of these states were computed. In the upper row, the down-(up-)facing triangles, connected by dashed lines, mark the highest (lowest)-lying ground state energies of these ten runs
at $1/N=1/25, 1/50, 1/60, 1/70, 1/80$ and $1/100$. In the lower panels, the $\langle\mathbf{L}^2_{\text{tot}}\rangle$ values are shown in the same way.
The energy ($\langle\mathbf{L}^2_{\text{tot}}\rangle$ values) obtained by combining the bases of the 10 individual runs into a single basis of $10\times N$ states is shown
as a solid line. From the relation between energy and $1/N$, we estimate the energy at $N=\infty$ (diamond marker) by extrapolation 
 (dotted line). The difference between the obtained extrapolation and the combined energy of $10\times 100=1000$ basis states is given by $0.0079 E^\infty_{2B}$ ($\alpha=1$), $0.0098 E^\infty_{2B}$ ($\alpha=2$), and $0.0121 E^\infty_{2B}$ ($\alpha=3$),
 and it can be regarded as an estimated basis set extrapolation error.
}
\label{ConvergenceR10fig2}
\end{figure*}
\begin{figure*}[t!]
\centering
\includegraphics[width=\linewidth]{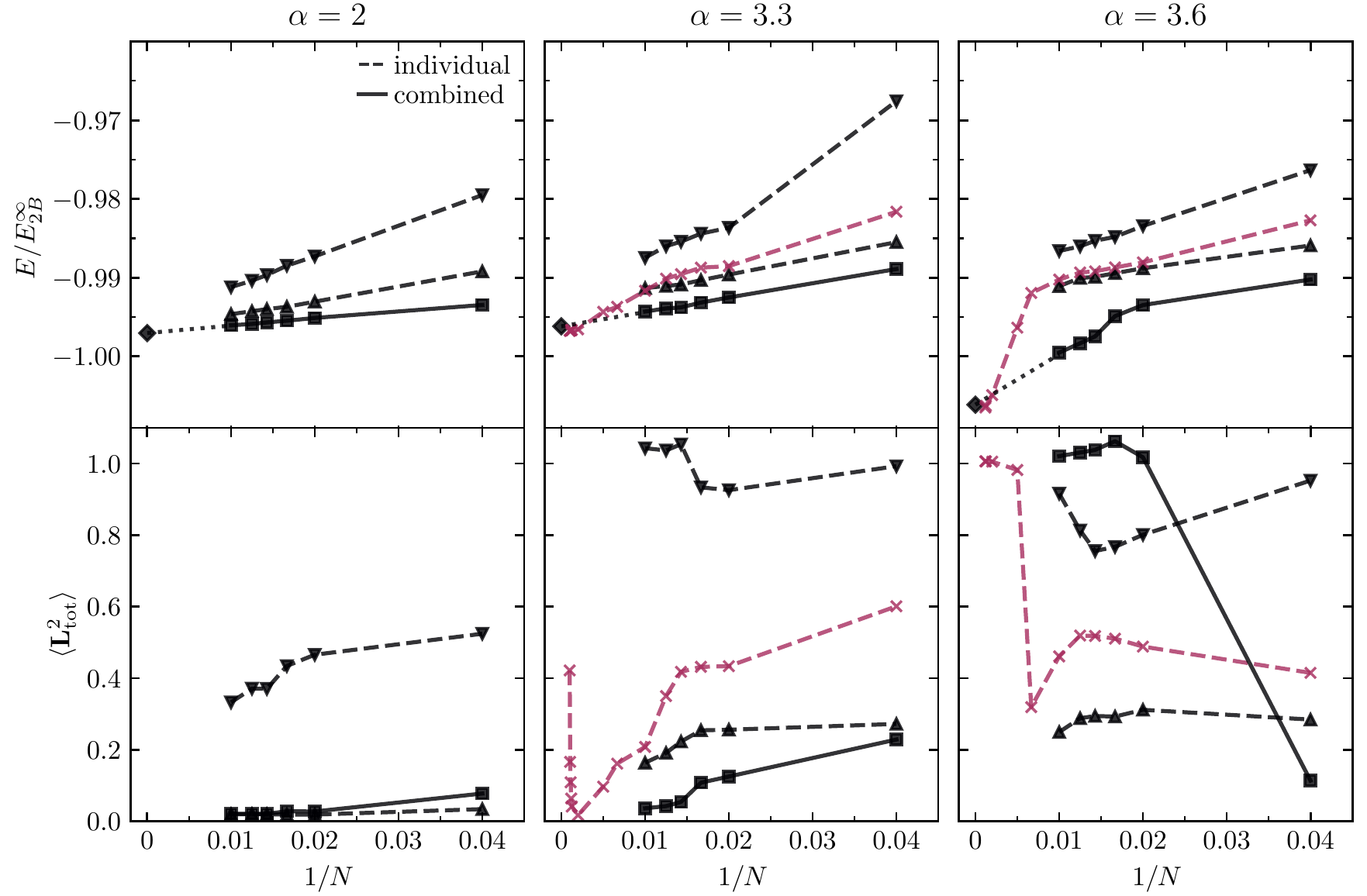}
\caption{Analysis of convergence with increasing number of basis states $N$, for parameters $R/r_B=100$, $r_0/r_B=0.2$ that yield a critical mass ratio of around $\alpha_c=3.3$ (see also \cref{fig:L_ext}). In the same manner as the analysis shown in \cref{ConvergenceR10fig2}, the analysis is conducted for mass ratios well before (left, $\alpha=2$), close to (middle, $\alpha=3.3$) and well after (right, $\alpha=3.6$) the dimer-to-trimer transition. Additionally, for $\alpha=3.3$ and $\alpha=3.6$, energies and angular momentum expectation values obtained in a single   run with up to $N=1000$ basis states are shown (crosses, dashed, purple). The difference between the 
combined energy of $10\times 100=1000$ basis states and the extrapolation is given by $0.00098 E^\infty_{2B}$ ($\alpha=2$), $0.0019 E^\infty_{2B}$ ($\alpha=3.3$) and $0.0066 E^\infty_{2B}$ ($\alpha=3.6$), and it can be regarded as an estimated basis set extrapolation error. The results obtained from the extrapolated energy (diamond) and the  single run with $N=1000$ basis states are consistent.}
\label{ConvergenceR100fig2}
\end{figure*}

\begin{figure*}[t!]
\centering
\includegraphics[width=\linewidth]{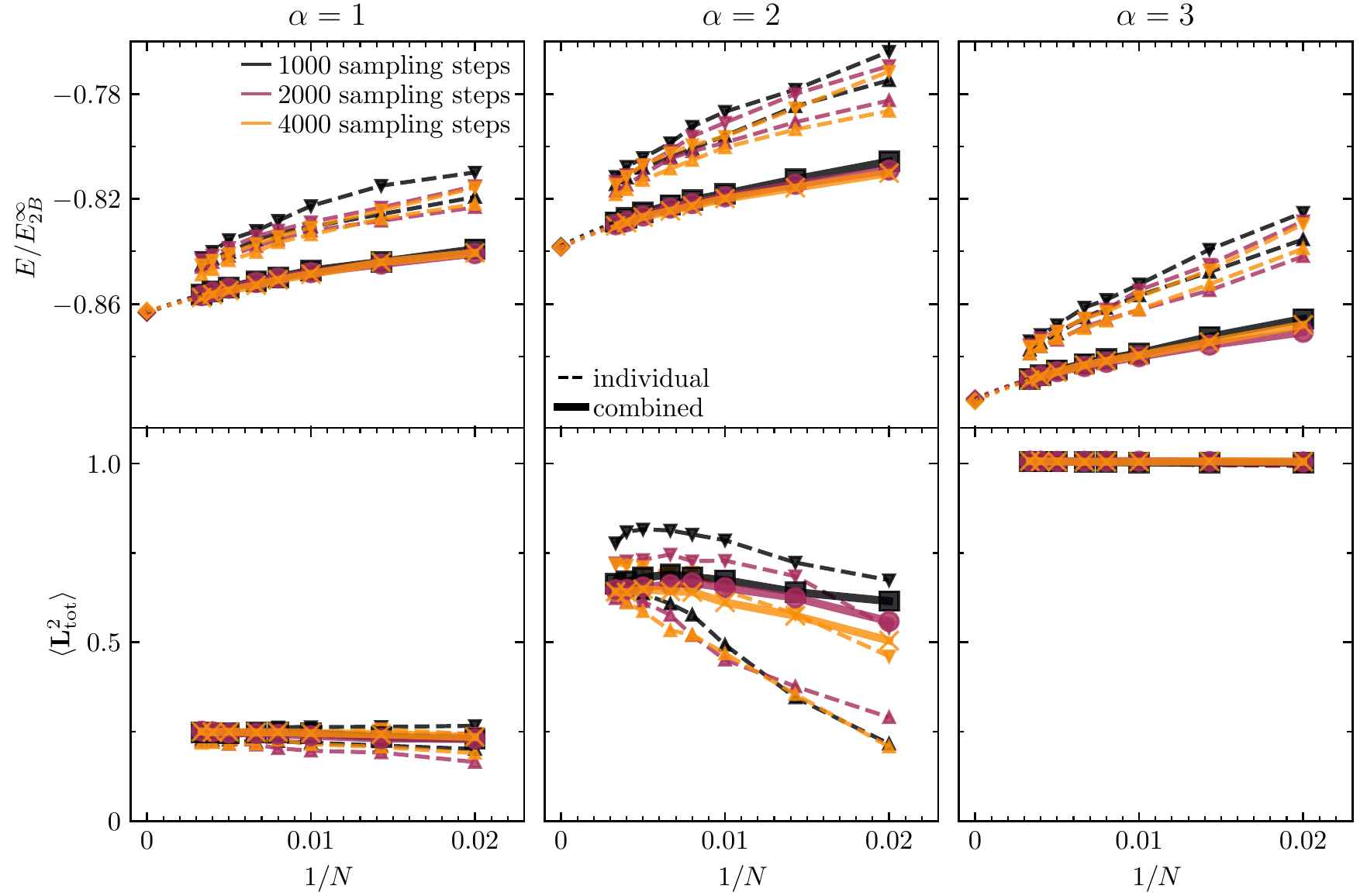}
\caption{Analysis of convergence as function of the number of basis states $N$ and increasing numbers of sampling steps. The analysis is conducted for the parameters  $R/r_B=10$, $r_0/r_B=0.8$ and for mass ratios well before (left, $\alpha=1$), close to (middle, $\alpha=2$) and well after (right, $\alpha=3$) the transition region. 10 independent runs with up to $300$ basis states using 1000 (black), 2000 (purple) and 4000 (yellow) sampling steps in every basis expansion step were conducted and the resulting ground states were tracked as a function of the number of basis states $N$ for $1/N= 1/300,1/250,1/200,1/150,1/125,1/100,1/70$ and $1/50$. Ground state energies and angular momentum expectation values are shown in the same manner as in \cref{ConvergenceR10fig2,ConvergenceR100fig2}, with different colours representing different numbers of sampling steps.}
\label{ConvergenceR10}
\end{figure*}

\begin{figure*}[t!]
\centering
\includegraphics[width=\linewidth]{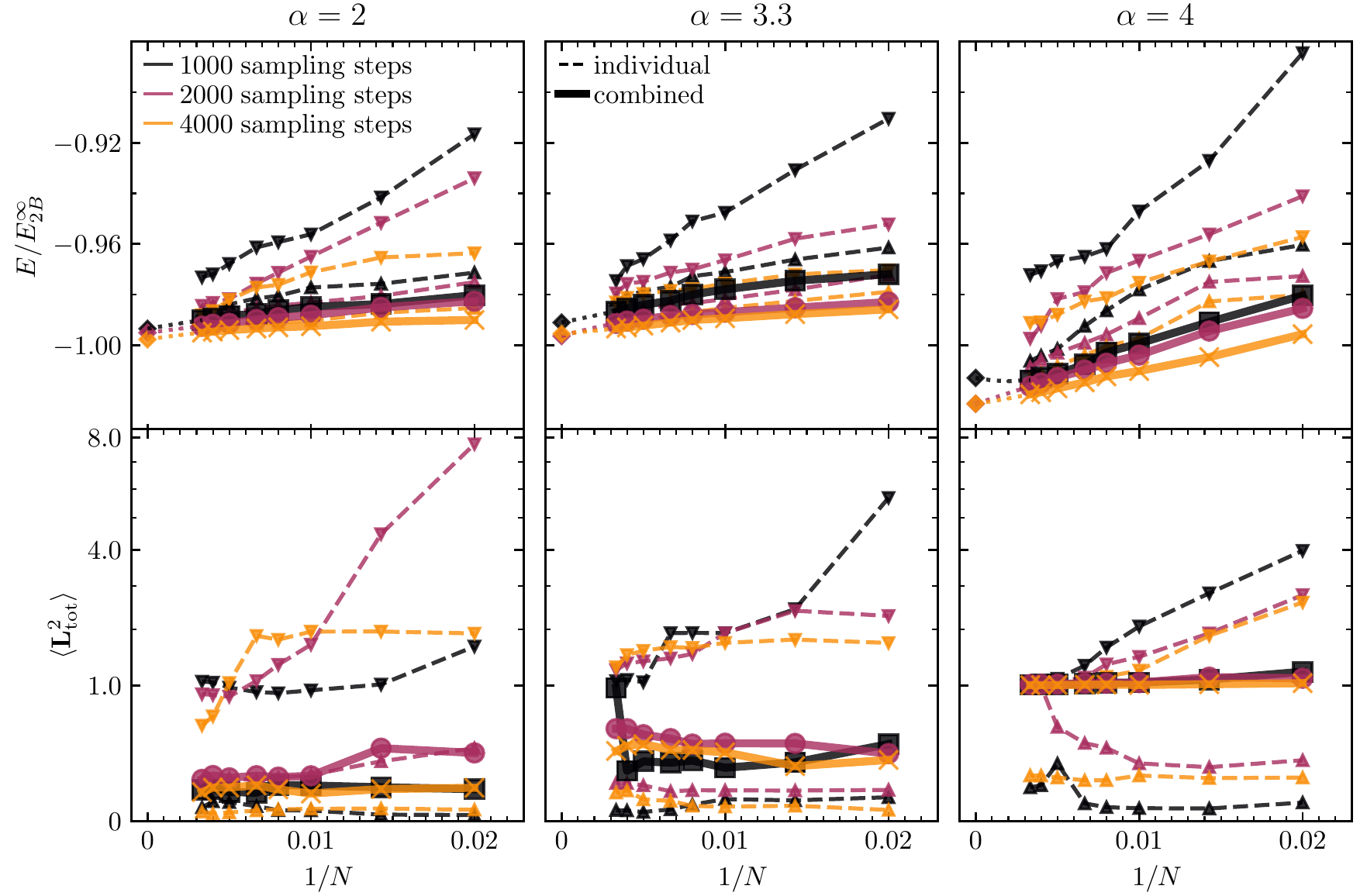}
\caption{Analysis of convergence with increasing number of basis states $N$ for $R/r_B=100$, $r_0/r_B=0.2$, and different numbers of sampling steps. In the same manner as the analysis shown in \cref{ConvergenceR10}, the analysis is conducted for mass ratios well before (left,$\alpha=2$), close to (middle,$\alpha=3.3$), and well after (right, $\alpha=4$) the transition region.}
\label{ConvergencerR100}
\end{figure*}

There are two aspects in which the SVM algorithm needs to achieve convergence in: 
\begin{enumerate}
    \item The number of basis states needs to be sufficiently large to describe the ground state accurately.
    \item A sufficient number of samples have to be drawn from the ECG manifold in every basis expansion step in order to ensure stable results. 
\end{enumerate}

The number of basis states and the number of samplings necessary to obtain accurate results varies depending on the nature of the ground state and the energy gap to the first excited state. Additionally, there are ranges of $R/r_B$ and $r_0/r_B$ that are more challenging to achieve convergence in. That is, when the confinement length $R$ is large and the interaction range $r_0$ is small, the manifold of wave functions that respect the confinement-imposed boundary conditions increases in size. In contrast, the subset of wave functions resolving the box potential is quite small. Combining both arguments, one sees that  a larger number of sampling steps is required. Additionally, when the energy gap between the dimer and the trimer state becomes small near the transition, the numerically-determined ground state can be a varying admixture of dimer and trimer state, resulting in angular momentum scatter. 

To study the convergence of the results shown in \cref{energyangular}, we have performed a convergence analysis of select data points. The results are shown in  \cref{ConvergenceR10fig2,ConvergenceR100fig2}, and they serve to investigate the behaviour of the energy and the angular momentum of the ground state as the number of basis states $N$ is increased. To further study the role of the number of sampling steps,  a similar analysis was performed in which, for varying numbers of sampling steps, the ground state properties were tracked, again, as a function of the number of basis states $N$. These results are shown in \cref{ConvergenceR10,ConvergencerR100}.

In \cref{ConvergenceR10fig2}, values of $R/r_B=10$, $r_0/r_B=0.8$ were chosen as representing  parameters for which it is easier to achieve 
convergence. In contrast, the values  $R/r_B=100$, $r_0/r_B=0.2$ chosen for  \cref{ConvergenceR100fig2} represent  parameters more challenging for the algorithm. For each of these sets of parameters, mass ratios before the transition, in the transition region, and beyond the transition were chosen to show the effect of the closing energy gap between the trimer and dimer states. A detailed description of the  data presented in the figures can be found in the respective figure captions.

As mandated by the variational principle, the  energies found are upper bounds for the true energy of the ground state. Moreover, as the number of basis states is increased, the variational energy must be lowered. In \cref{ConvergenceR10fig2}, it can be seen that away from the transition ($\alpha=1$, $\alpha=3$)
the 10 individual energies and angular momenta have a tight grouping, indicating that the number of sampling steps is sufficient. 
At the transition, the individual energies are also grouped tightly, but because the energy gap to the first excited state is small, the angular momentum expectation values have a significant spread and a stabilization of the observable comes from the combination of individual runs.  \cref{ConvergenceR100fig2}, on the other hand, is obtained for a larger system size, making the dimer-to-trimer crossover much more narrow. As a consequence, the spread of energies relative to the energy gap to $-E_{2B}^\infty$ is much larger than in \cref{ConvergenceR10fig2}. Highlighting the challenge in describing such parameter regimes, even away from the transition, stabilization of the results is achieved only after the combination of basis states of the individual runs, and not by a sheer increase of the number of sampling 
steps as in \cref{ConvergenceR10fig2}.   

Away from the transition, no qualitative changes in the angular momentum expectation value are observed once around 50 states have been taken into account. This holds true even when comparing with a run in which the basis states are derived from a single run of up to 1000 basis states (see also \cref{fig:L_ext}), rather than from 10 independent runs of up to 100 basis states. Close to the transition, however, a larger number of basis states is required to achieve convergence, as both the ground and the first excited state need to be resolved with a sufficiently large number of basis states. As a result, the single run of up to 1000 basis states shows  different results than the combined runs at $\alpha=3.3$ shown in \cref{ConvergenceR100fig2}, see also the discussion in \cref{app:convergence_analysis_a}.

\begin{table*}
\centering
\begin{tabular}{|x{3cm}|x{1.2cm}|x{1.2cm}|x{1.2cm}|x{1.2cm}|x{1.2cm}|x{1.2cm}|}
\multicolumn{7}{c}{extrapolation error $\left[ E_{2B}^{\infty} \right]$}\\
\hline
\multirow{2}*{}&\multicolumn{3}{|c|}{$R/r_B=10,\;r_0/r_B=0.8$}&\multicolumn{3}{|c|}{$R/r_B=100,r_0/r_B=0.2$}\\
\cline{2-7}
&$\alpha=1$&$\alpha=2$&$\alpha=3$&$\alpha=2$&$\alpha=3.3$&$\alpha=4$\cr
\hline
1000 sampling steps&0.0068&0.0092&0.0079&0.0033&0.0043&0.0009\cr
\hline
2000 sampling steps&0.0065&0.0087&0.0072&0.0026&0.0050&0.0069\cr
\hline
4000 sampling steps&0.0051&0.0081&0.0081&0.0026&0.0021&0.0037\cr
\hline
\end{tabular}
\caption{Estimated extrapolation error of the ground state energy in units of $ E_{2B}^{\infty}$ in dependence of the number of sampling steps, for different system parameters. The uncertainty is obtained by comparing the extrapolated energy and the  energy obtained by a combination of $10\times100$ basis states, shown in \cref{ConvergenceR10} and \cref{ConvergencerR100}.} 
\label{TableDifference}
\end{table*}

For the data shown in \cref{ConvergenceR10fig2,ConvergenceR100fig2}  we estimate the uncertainties of our energies as the energy difference between the combined energies at $N=100$, and the extrapolated energies at $1/N=0$. The resulting uncertainties are given in the captions of \cref{ConvergenceR10fig2,ConvergenceR100fig2}. We note that these estimated uncertainties are of the order of $\sim 0.01 E_{2B}^\infty$ in \cref{ConvergenceR10fig2}, and, in \cref{ConvergenceR100fig2}, they are of the order $\sim 0.001   E_{2B}^\infty$ before the transition and about  $\sim 0.007   E_{2B}^\infty$ beyond the transition. As such, they are much smaller than the actual gap between the ground state 
 and $-E_{2B}^\infty$.  Furthermore, we  note that,  as can be seen in \cref{ConvergenceR100fig2},  the extrapolated energies are very close to the energies obtained from a single run of 1000 basis states.

Finally, we investigate the impact of the number of sampling steps on convergence. To this end, we show in \cref{ConvergenceR10}  an analysis of the convergence with the number of basis states for relatively low numbers of sampling steps obtained for $R/r_B=10$ and $r_0/r_B=0.8$. In \cref{ConvergencerR100}, the same analysis is performed for $R/r_B=100$ and $r_0/r_B=0.2$. In the case of $R/r_B=10$ and $r_0/r_B=0.8$, the energies and angular momenta are, along with their spreads, comparable to those shown in \cref{ConvergenceR10fig2}, even though the former results were obtained  for a significantly lower number of sampling steps. In contrast, in the case of $R/r_B=100$ and $r_0/r_B=0.2$, shown in \cref{ConvergencerR100}, one can see that, by increasing the number of sampling steps, one obtains a much tighter grouping in energy, which differs from the results shown in \cref{ConvergenceR100fig2}. The estimated uncertainties obtained from  the data given in \cref{ConvergenceR10,ConvergencerR100} are shown in
\cref{TableDifference}. 
This illustrates further the requirements different parameter ranges of $r_0/r_B$ and $R/r_B$ pose on the number of sampling steps.

\clearpage

\section{REDUCED DENSITY DISTRIBUTION WITH $\theta$ AND $R_2$}\label{apptheta}
\begin{figure}[b!]
\centering
\includegraphics[width=\linewidth]{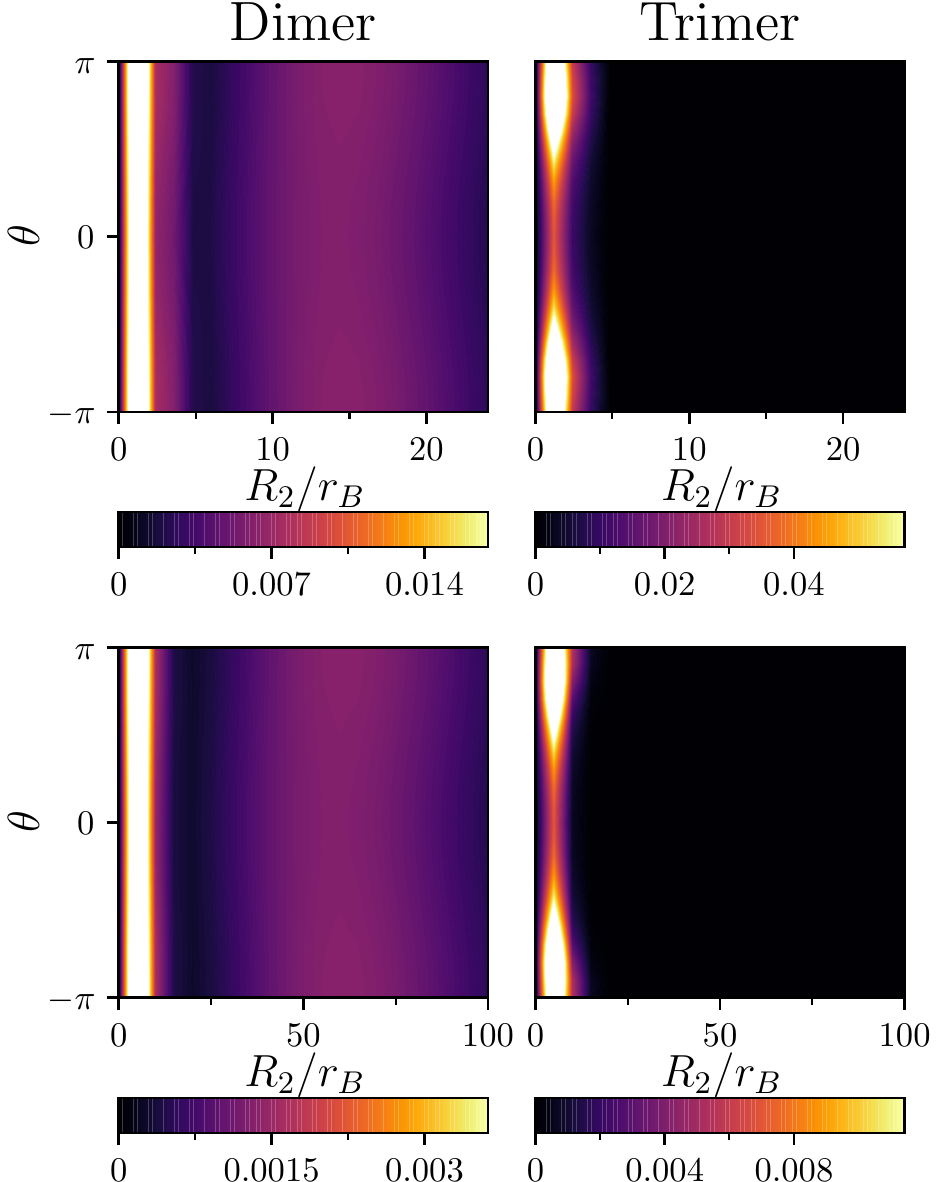}
\caption{Reduced density distribution $u_3(R_2,\theta)r_B$ for systems with and without Coulomb interaction. Upper panel: Results for a dimer ($\alpha=2$, left) and a trimer state ($\alpha=3$, right)  for $r_0/r_B=0.8$, $R/r_B=20$,  and $q=0$. Lower panel: Results for a dimer ($\alpha=2$, left) and a trimer state ($\alpha=3.5$, right)  for $r_0/r_B=0.8$, $R/r_B=100$ and $q=0.3$. For the trimer state, $u_3(R_2,\theta)r_B$ has an angular dependence such that it achieves its minimal value around $\theta=0$ and its maximal value around $\theta=\pi$. This shows that the fermions in the trimer state have a preference for an anti-parallel configuration that is mainly caused by Pauli exclusion.}\label{u3function}
\end{figure}

To further study the anatomy of the dimer and trimer states with respect to their angular distribution, we define the reduced density distribution $u_3(R_2,\theta)$ as
\begin{equation}
    u_3(R_2,\theta)=R_2\int d^2\mathbf{r}_1d^2\mathbf{R}_3 |\Psi(\mathbf{r}_1,\mathbf{r}_1+\mathbf{R}_2,\mathbf{r}_1+\mathbf{R}_3)|^2.
\end{equation}
Here, the vectors $\mathbf{R}_2$ and $\mathbf{R}_3$ are parametrized as $\mathbf{R}_2=R_2(\cos{(\theta_3+\theta)},\sin{(\theta_3+\theta)})$, $\mathbf{R}_3=R_3(\cos{\theta_3},\sin{\theta_3})$. In \cref{u3function}, we show the density distribution $u_3(R_2,\theta)$ for  dimer and a trimer states. For the trimer states, when $R_2$ is close to 0, the density distribution $u_3$ almost vanishes around $\theta=0$ and achieves its maximum at approximately $\theta=\pi$ which shows that the fermions tend to locate at opposite sides of the impurity mainly due to Pauli exclusion.  On the other hand, $u_3$ shows no visible angular dependence for the dimer state as the distance between the two fermions is relatively large and thus Pauli exclusion does not play an important role.
\clearpage
\end{document}